\documentclass[twocolumn,twocolappendix]{aastex631}
\let\splitbox\relax

\usepackage{xspace}
\usepackage{hyperref}
\usepackage{enumitem}
\usepackage{amssymb}
\usepackage{amsmath}
\usepackage{pifont}

\graphicspath{{./}{figures/}}
\usepackage{lipsum}
\usepackage{graphicx}
\usepackage{subfigure}
\usepackage{comment}
\usepackage{booktabs}
\usepackage{amsmath}
\usepackage{etoolbox}
\usepackage{IEEEtrantools}
\usepackage{parskip}
\usepackage{ragged2e}
\usepackage{textcomp,gensymb}
\usepackage{physics}
\usepackage{multirow}
\let\tablenum\relax
\usepackage{siunitx} % For alignment of numbers
\usepackage{adjustbox}

\usepackage{tabularx, booktabs,array}
\begin{document}
\UseRawInputEncoding
%\linenumbers
%\switchlinenumbers

\title{Lensing-Reconstructed Dark Matter-Intracluster Medium Coherence as a Probe of Cluster Dynamical State: Application to HSTFF, RELICS, and CLASH Clusters}
\footnotetext{\copyright\ 2026. All rights reserved.}

\correspondingauthor{Giulia Cerini}
\email{giulia.cerini@jpl.nasa.gov}

\author[0000-0002-5273-4634]{Giulia Cerini}
\affiliation{Jet Propulsion Laboratory, California Institute of Technology, 4800 Oak Grove Dr, Pasadena, CA 91109, USA}

\author[0000-0002-6044-2164]{Sayan Saha}
\affiliation{Northeastern University, 360 Huntington Ave, Boston, MA 02115}

\author[0000-0002-9883-7460]{Jacqueline McCleary}
\affiliation{Northeastern University, 360 Huntington Ave, Boston, MA 02115}

\author[0009-0003-9547-0952]{Eric Habjan}
\affiliation{Northeastern University, 360 Huntington Ave, Boston, MA 02115}

\author[0000-0002-1697-186X]{Nico Cappelluti} 
\affiliation{Department of Physics, University of Miami, 1320 S Dixie Hway, Coral Gables, FL 33146, USA}

\author[0000-0002-5554-8896]{Priyamvada Natarajan}
\affiliation{Department of Astronomy, Yale University, 219 Prospect Street, New Haven, CT 06511, USA}
\affiliation{Department of Physics, Yale University, P.O. Box 208121, New Haven, CT 06520, USA}
\affiliation{Black Hole Initiative, Harvard University, 20 Garden Street, Cambridge, MA 02138, USA}

\author{Sabina Khizroev}
\affiliation{Duke University, 2080 Duke University Road, Durham, NC 27708, USA}

\author[0000-0002-4485-8549]{Jason Rhodes}
\affiliation{Jet Propulsion Laboratory, California Institute of Technology, 4800 Oak Grove Dr, Pasadena, CA 91109, USA}

\author[ 	
0000-0002-9378-3424]{Eric Huff}
\affiliation{Jet Propulsion Laboratory, California Institute of Technology, 4800 Oak Grove Dr, Pasadena, CA 91109, USA}

\author{Nicole Chidester}
\affiliation{Northeastern University, 360 Huntington Ave, Boston, MA 02115}

\author{Maya Amit}
\affiliation{New York University, 50 West 4th Street New York, NY, 10012}

\author[0000-0002-0086-0524]{Andrew Robertson}
\affiliation{Carnegie Observatories, 813 Santa Barbara St, Pasadena, CA 91101}

\author[0000-0001-6928-4345]{Bryanne McDonough}
\affiliation{Adelphi University, 1 South Ave, Garden City, NY 11530}

\author[0000-0001-6411-3686]{Elena Bellomi}
\affiliation{Center for Astrophysics Harvard $\&$ Smithsonian, 60 Garden St, Cambridge, MA 02138}

\author[0000-0001-8914-8885]{Erwin T. Lau}
\affiliation{Nara Womens University, Kitauoyanishi-machi, Nara, Nara 630-8506, Japan}

\author[0000-0003-3175-2347]{John ZuHone}
\affiliation{Center for Astrophysics Harvard $\&$ Smithsonian, 60 Garden St, Cambridge, MA 02138}

%\author{John ZuHone}
%\affiliation{Center for Astrophysics $\vert$ Harvard $\&$ Smithsonian, 60 Garden St., Cambridge, MA 02138, USA}

\begin{abstract} \label{abstract}
We present the first application of Fourier-space coherence analysis between the lensing-reconstructed projected mass distribution and the X-ray-emitting intracluster medium to a sample of 49 observed galaxy clusters. Using publicly available HST convergence maps from the Hubble Frontier Fields, CLASH, and RELICS programs, together with Chandra X-ray imaging, we measure the scale-dependent coherence between the dark-matter-dominated surface mass density and the hot baryonic gas. We use the coherence length, $\ell_{\rm CR}$, defined as the scale above which the two maps remain at least $90\%$ coherent, as a diagnostic of cluster dynamical state. Across the sample, dynamically relaxed systems exhibit high coherence over a broad range of scales and small $\ell_{\rm CR}/r_{500}$, while disturbed and merging systems show a loss of coherence on intermediate and small scales, yielding larger $\ell_{\rm CR}/r_{500}$. The inferred coherence lengths show sensitivity to lens-model assumptions and to the heterogeneous extent of the available convergence maps. Nevertheless, the coherence signal remains physically interpretable and provides a stringent measure of dark-matter-gas alignment. Applying a conservative threshold, $\ell_{\rm CR}/r_{500}<0.2$, we find that only $\sim16\%$ of the sample is relaxed; this fraction rises to $\sim41\%$ for a more permissive threshold of $\ell_{\rm CR}/r_{500}<0.4$. Relative to previous X-ray and morphological classifications, we find a $\sim24\%$ disagreement, with the coherence method identifying more systems as dynamically disturbed. These results demonstrate that lensing-X-ray coherence provides a complementary, scale-resolved probe of cluster dynamical state, while highlighting the need for homogeneous, wide-field weak-lensing maps to control reconstruction and field-of-view systematics.

\end{abstract}


\section{Introduction} \label{sec:intro}
Galaxy clusters, the most massive gravitationally bound systems in the Universe, with total masses in the range of $10^{14}-10^{15}\ M_{\odot}$, consist of dark matter (DM) halos filled with X-ray emitting hot gas with typical temperatures of $10^{7}-10^{8}\ \mathrm{K}$ - the intracluster medium (ICM) - and hundreds to thousands of galaxies. The wide variety of physical processes operating within these environments makes clusters fundamental probes of both galaxy evolution and cosmology. 

 \begin{figure*}
\centering
      \subfigure[]{\includegraphics[width=0.33\textwidth]{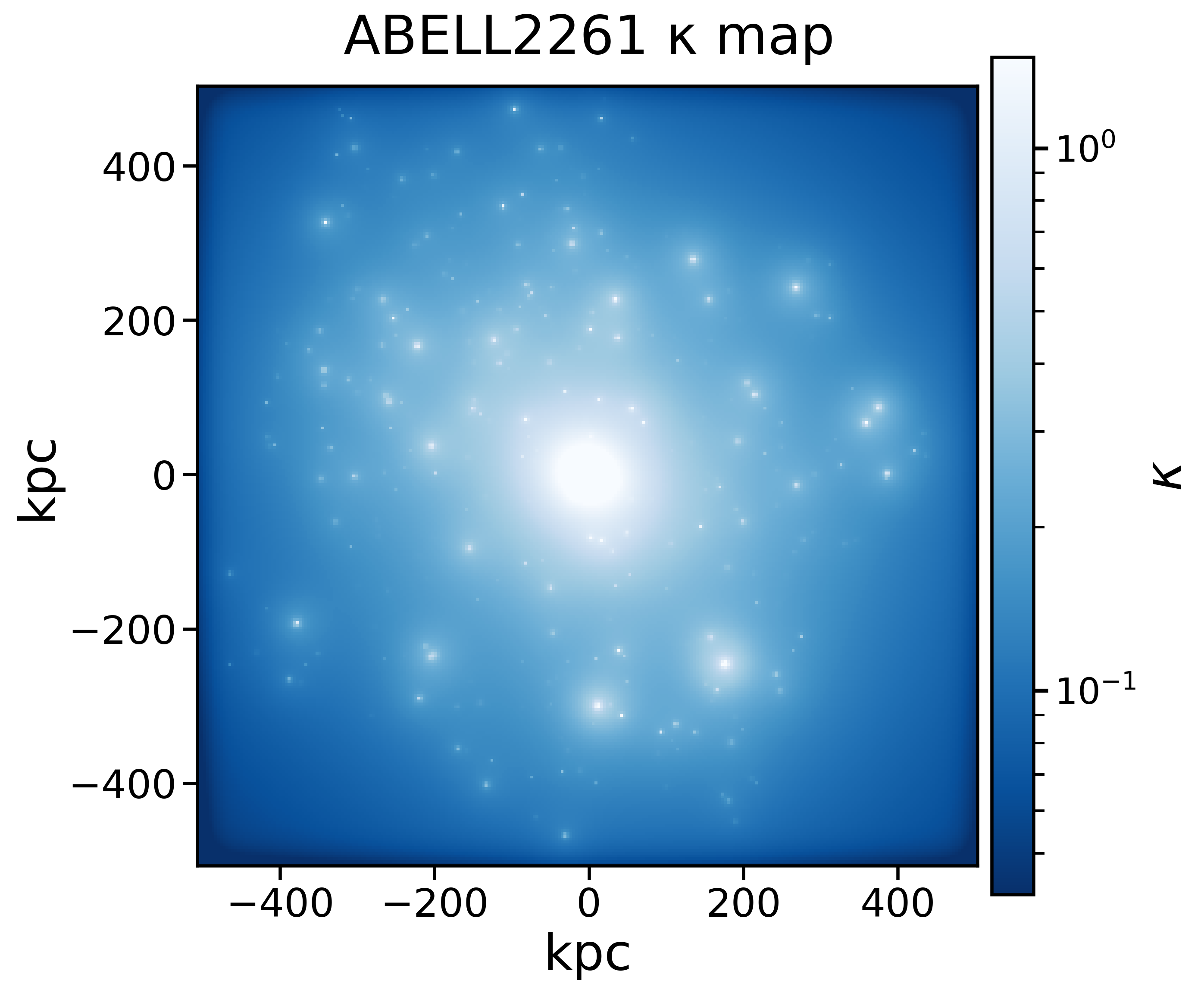}}
      \subfigure[]{\includegraphics[width=0.33\textwidth]{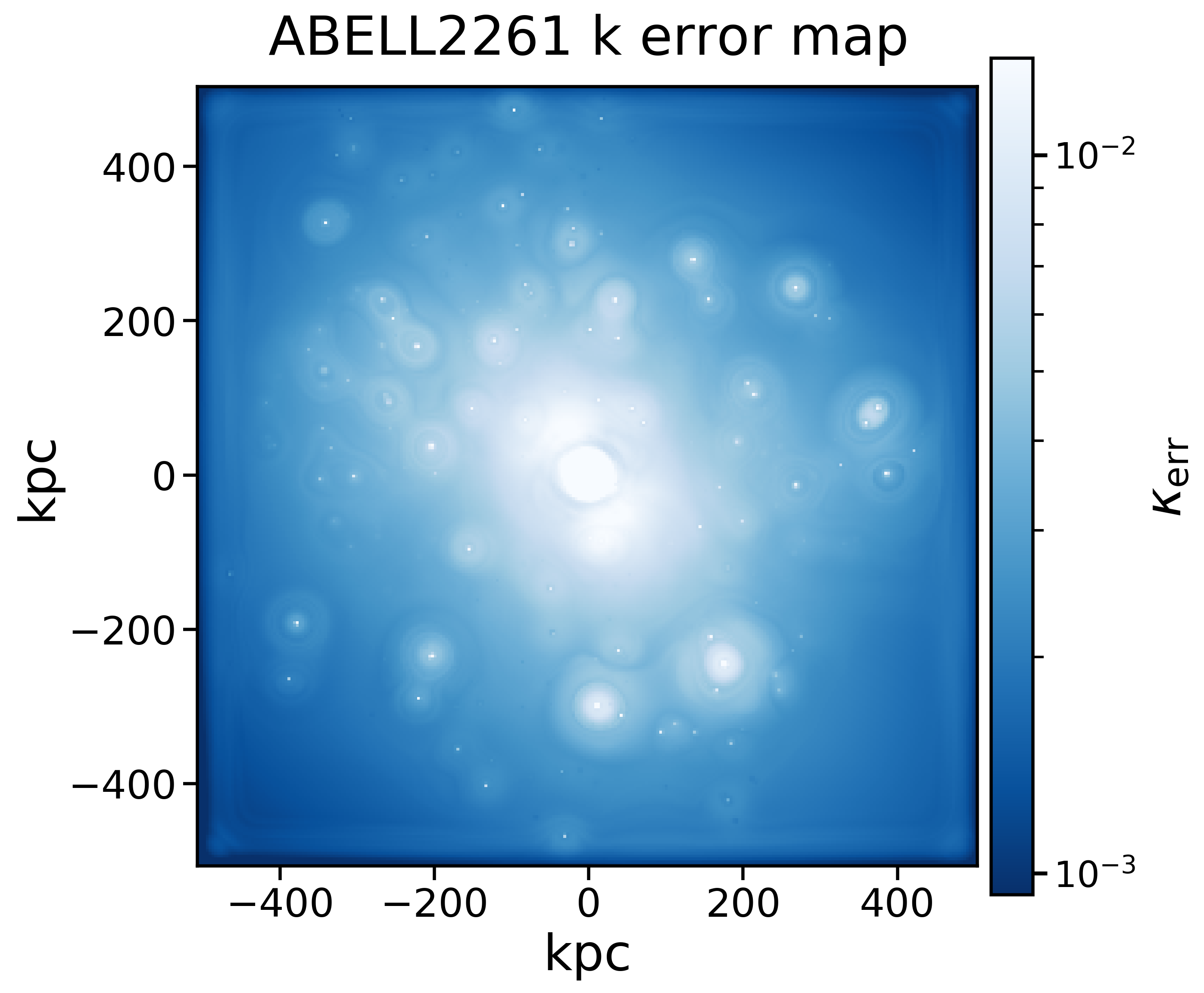}}
      \subfigure[]{\includegraphics[width=0.28\textwidth]{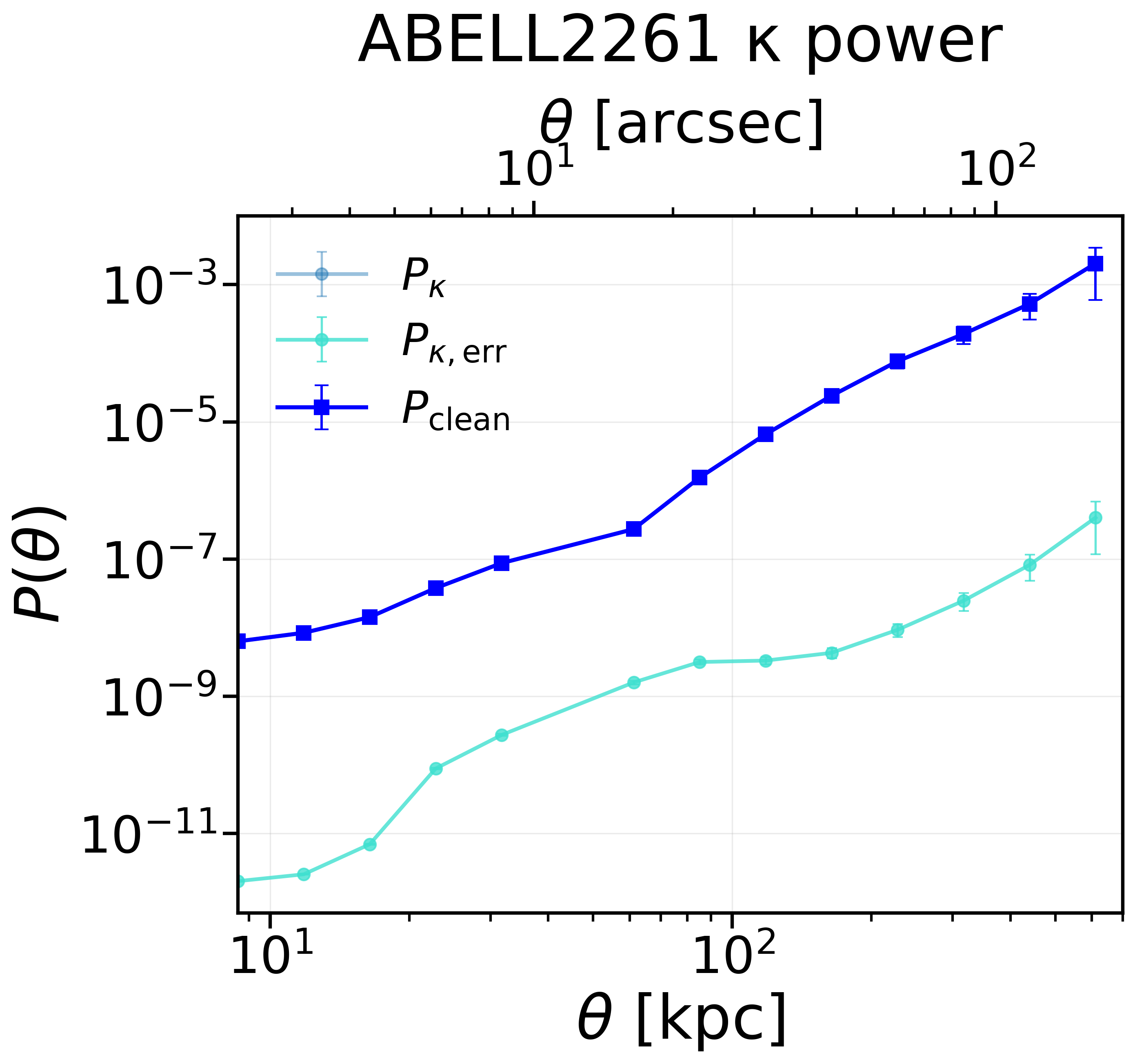}}     
    \caption{Left panel: Convergence map $\kappa(\boldsymbol{x})$ of the galaxy cluster ABELL2261 reconstructed with the \texttt{zitrin\_ltm\_gauss\_v2} model. Central panel: Corresponding $\kappa(\boldsymbol{x})$ uncertainty map. Right panel: Power spectrum of the convergence map (light blue), of the associated uncertainty map (turquoise), and of the cleaned signal (blue), obtained by subtracting the noise contribution from the raw power spectrum.}
      
 \label{kappa_example}
 \end{figure*}

  \begin{figure*}
\centering
      \subfigure[]{\includegraphics[width=0.32\textwidth]{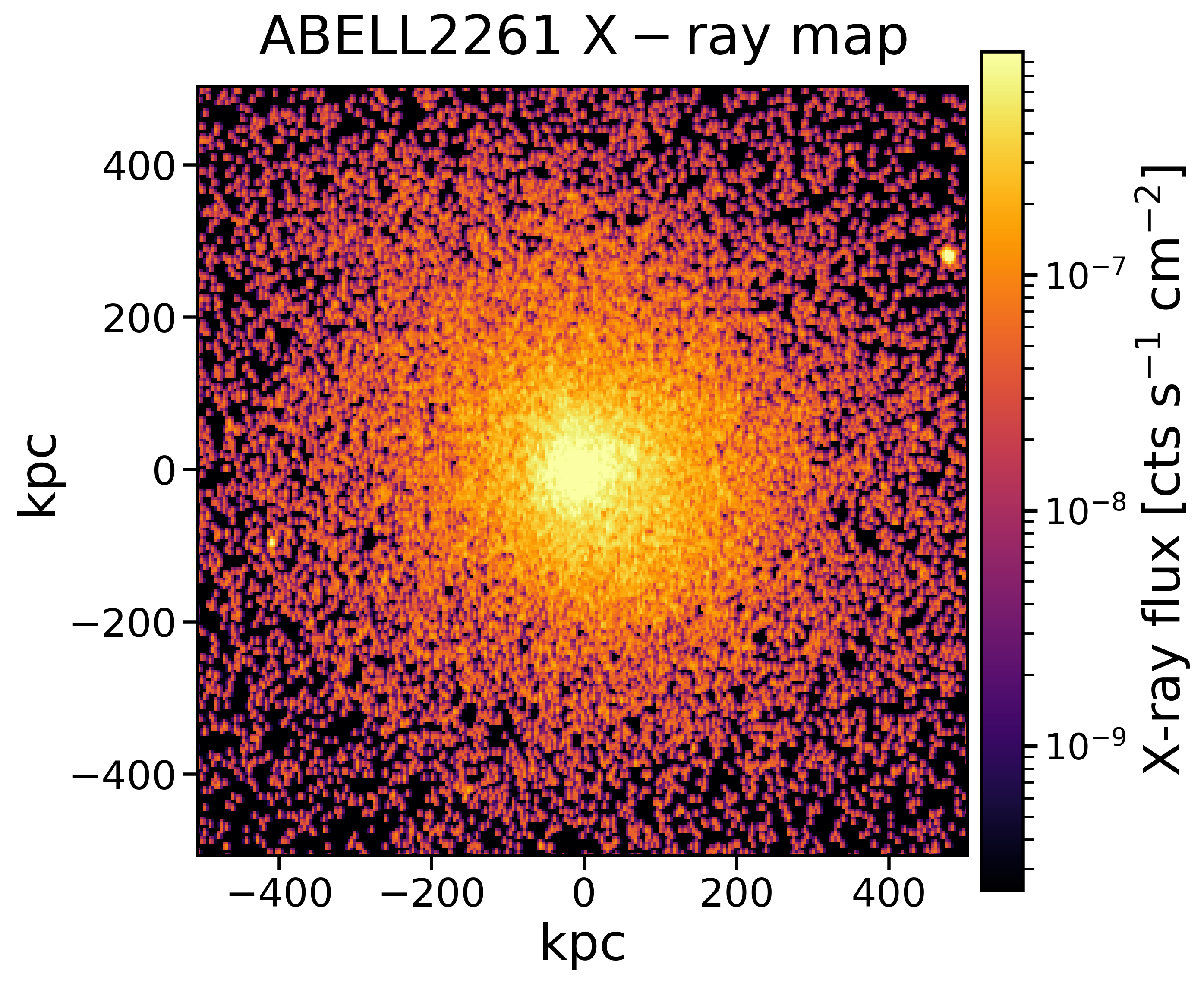}}
      \subfigure[]{\includegraphics[width=0.335\textwidth]{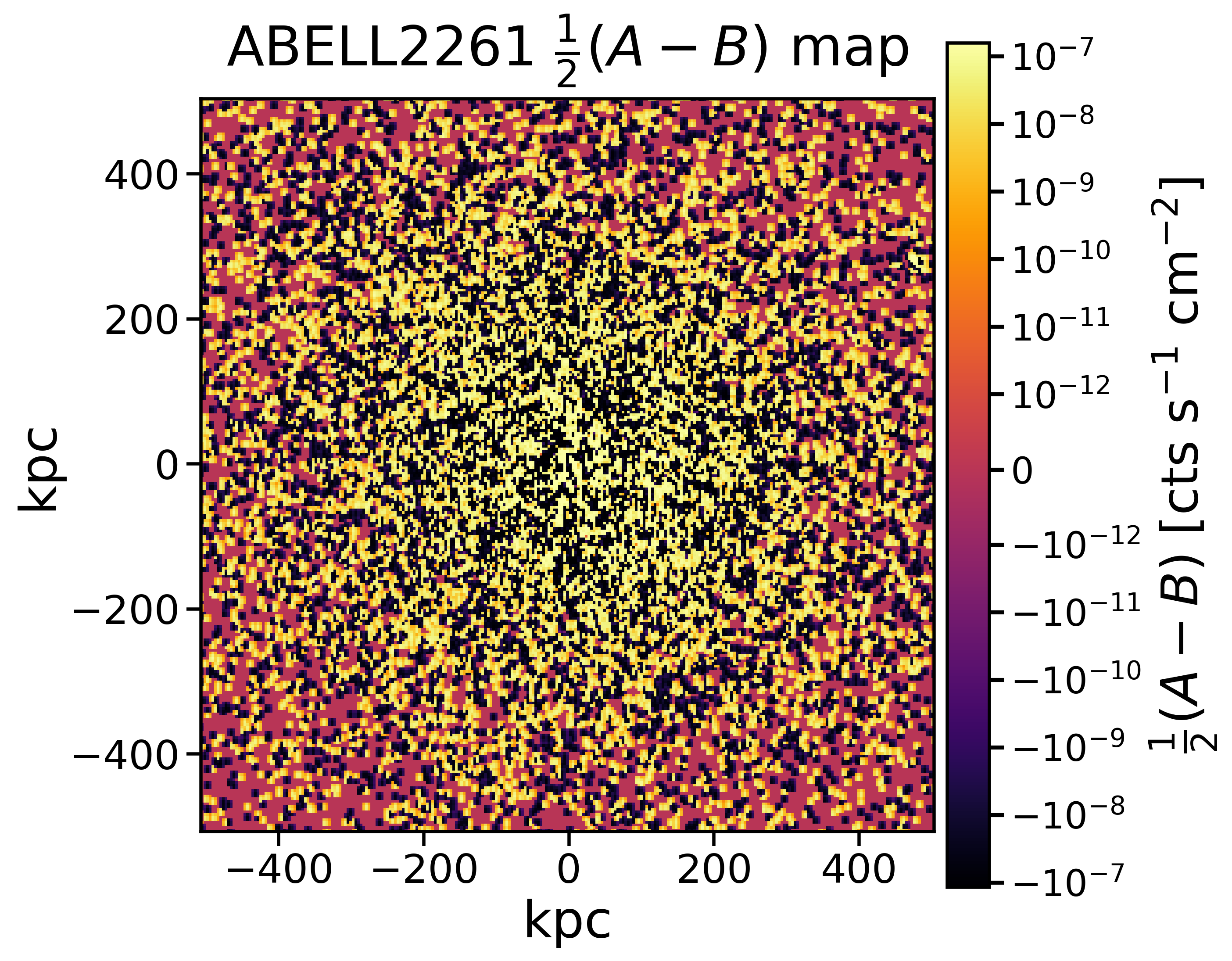}}
      \subfigure[]{\includegraphics[width=0.27\textwidth]{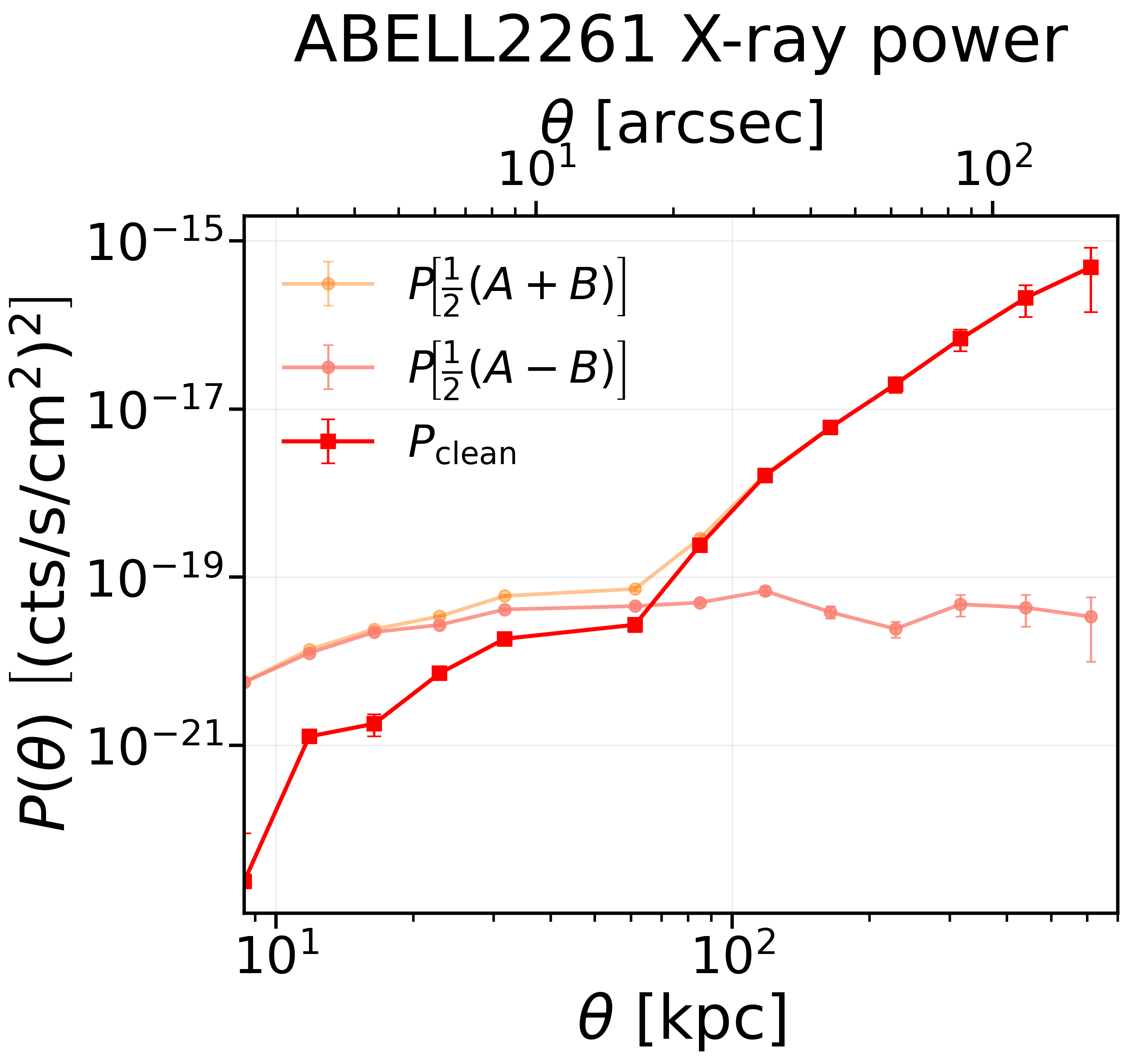}}     
    \caption{Left panel: background-subtracted, exposure-corrected X-ray $\tfrac{1}{2}(A+B)$ map of the galaxy cluster ABELL2261. 
Central panel: exposure-corrected X-ray $\tfrac{1}{2}(A-B)$ map of ABELL2261. 
Right panel: corresponding power spectra of the $\tfrac{1}{2}(A+B)$ map (orange), the $\tfrac{1}{2}(A-B)$ map (salmon), and the cleaned signal (red), obtained by subtracting the noise power spectrum from the total.}
      
 \label{xray_example}
 \end{figure*}

One of the fundamental properties of a galaxy cluster is its dynamical state, namely the degree to which the system is dynamically relaxed. A dynamically relaxed cluster has reached a state of approximate equilibrium between gravity (dominated primarily by DM) and internal pressure. Relaxed clusters are typically characterized by smooth dark matter and intracluster gas distributions, which closely trace each other, resulting in a regular overall mass distribution. They also exhibit a more spherical morphology and lack significant substructures or ongoing merger activity. The dynamical state plays a crucial role in many contexts, including constraining cosmological parameters and interpreting observables sensitive to the nature of DM. Improving the precision of cosmological parameter estimates is a central goal of modern cosmology. In particular, measurements of the amplitude of matter fluctuations, $\sigma_{8}$ (e.g., \citealt{Heymans2021}), and the Hubble constant, $H_{0}$ (e.g., \citealt{2021DiValentino}), remain in significant tension across different observational probes. Notably, more recent results from galaxy cluster cosmology based on the \textit{eROSITA} All-Sky Survey Data Release 1 (eRASS1; \citealt{ghirardini2024srgerosita}) and updated cosmic shear analyses from the final data release of the Kilo-Degree Survey (KiDS-Legacy; \citealt{Stolzner2025}) indicate a partial alleviation of the $\sigma_{8}$ tension, showing improved consistency with \textit{Planck} measurements (\citealt{Planck2024}). 

Cosmological constraints derived from galaxy cluster abundance rely critically on accurate mass estimates, which remain one of the major challenges in modern cosmology, even for nearby systems. Cluster mass is not directly observable and must be inferred from its gravitational and baryonic tracers. Common approaches include gravitational lensing (\citealt{Bonnet94,Fahlman_1994}), the Sunyaev–Zel'dovich (SZ) effect (\citealt{S1970,S1980}), X-ray emission (\citealt{Evrard1996}), and galaxy dynamics (\citealt{Merritt1987}). Among these methods, gravitational lensing provides the most direct probe of the total projected mass distribution, as it relies on the gravitational deflection of background light rather than assumptions about the dynamical state of the cluster. The latter three approaches, instead, assume that the ICM and galaxy distributions are in hydrostatic or virial equilibrium within the DM potential. However, this assumption often breaks down in clusters undergoing mergers or accretion events, where gas motions generate non-thermal pressure and drive deviations from equilibrium (e.g. \citealt{Lau_2009}, \citealt{Churazov_2012}; \citealt{zhuravleva17}; \citealt{https://Cerini22}). As a result, clusters that are dynamically unrelaxed can have their masses underestimated by up to $\sim20\%$ (\citealt{Nagai2007}, \citealt{Kravtsov2012}). Feedback processes, turbulence, and shocks can further disturb the gas, leaving observable imprints in temperature and morphology that violate the self-similar expectations of purely gravitational evolution (e.g. \citealt{Bhat2008}, \citealt{McCarthy2010}, \citealt{Fabjan2011}, \citealt{Bulbul_2019}). Nevertheless, gravitational lensing is also affected by projection effects and modeling uncertainties (e.g. \citealt{Jauzac+2015}, \citealt{Lee2023}). As a result, scaling relations based on these mass estimates continue to show deviations from the self-similar scenario. Altogether, these limitations highlight the cluster dynamical state as a key source of systematic uncertainty in cosmological analyses. In this context, a robust characterization of the dynamical state is essential to mitigate astrophysical sources of scatter and bias associated with departures from equilibrium, thereby improving the reliability of cosmological constraints.

Beyond cosmological parameters, the dynamical state also plays a key role in the interpretation of observables aimed at probing the nature of DM. Models of self-interacting DM (SIDM) have been proposed as an extension of the cold DM (CDM) paradigm, potentially alleviating small-scale tensions while preserving its large-scale success (\citealt{Spergel2000}). In this framework, DM particles can exchange momentum through elastic scattering, which may leave observable signatures in galaxy clusters. A promising way to detect such signatures is through measurements of spatial offsets between the DM distribution (inferred from gravitational lensing), the galaxy component and the hot intracluster gas traced by X-ray emission (e.g. \citealt{Markevitch2004}, \citealt{Clowe_2006}, \citealt{Randall2008}, \citealt{Bradac2008}, \citealt{Harvey2015}, \citealt{Robertson17}, \citealt{Sirks2024}). However, in disturbed systems, projection effects, substructures, and merger-induced asymmetries can bias the identification of component peaks, artificially enhancing or suppressing the measured offsets. Gas stripping during mergers may further introduce asymmetries in the dark matter and galaxy distributions, complicating the interpretation of spatial offsets between cluster components (\citealt{Robertson17}, \citealt{Wittman2018}). In contrast, more relaxed clusters are expected to exhibit a higher degree of alignment between components.

Throughout the literature, numerous observable properties of galaxy clusters have been used as indicators of their dynamical state, each aiming to quantify how relaxed a system is. Some of the most commonly used X-ray estimators include: the centroid shift of the X-ray surface brightness(\citealt{Mohr1993}); the power ratios, based on multipole moments of the X-ray surface brightness, which capture deviations from circular symmetry and reveal substructures (\citealt{Buote1995}); the cuspiness of the gas density profile, measuring how peaked the core is and tracing the presence of cool cores (\citealt{Vikhlinin2007}); the concentration ratio, defined as the ratio of X-ray flux within small and large apertures, indicating the central compactness of the emission (\citealt{Santos2008}); the central gas density (\citealt{Hudson2010}); and the Gini coefficient, which quantifies the inequality of pixel flux distribution and serves as a measure of morphological irregularity (\citealt{Parekh2015}). 

\begin{figure}[ht]
 \centering
 \includegraphics[width=0.8\columnwidth]{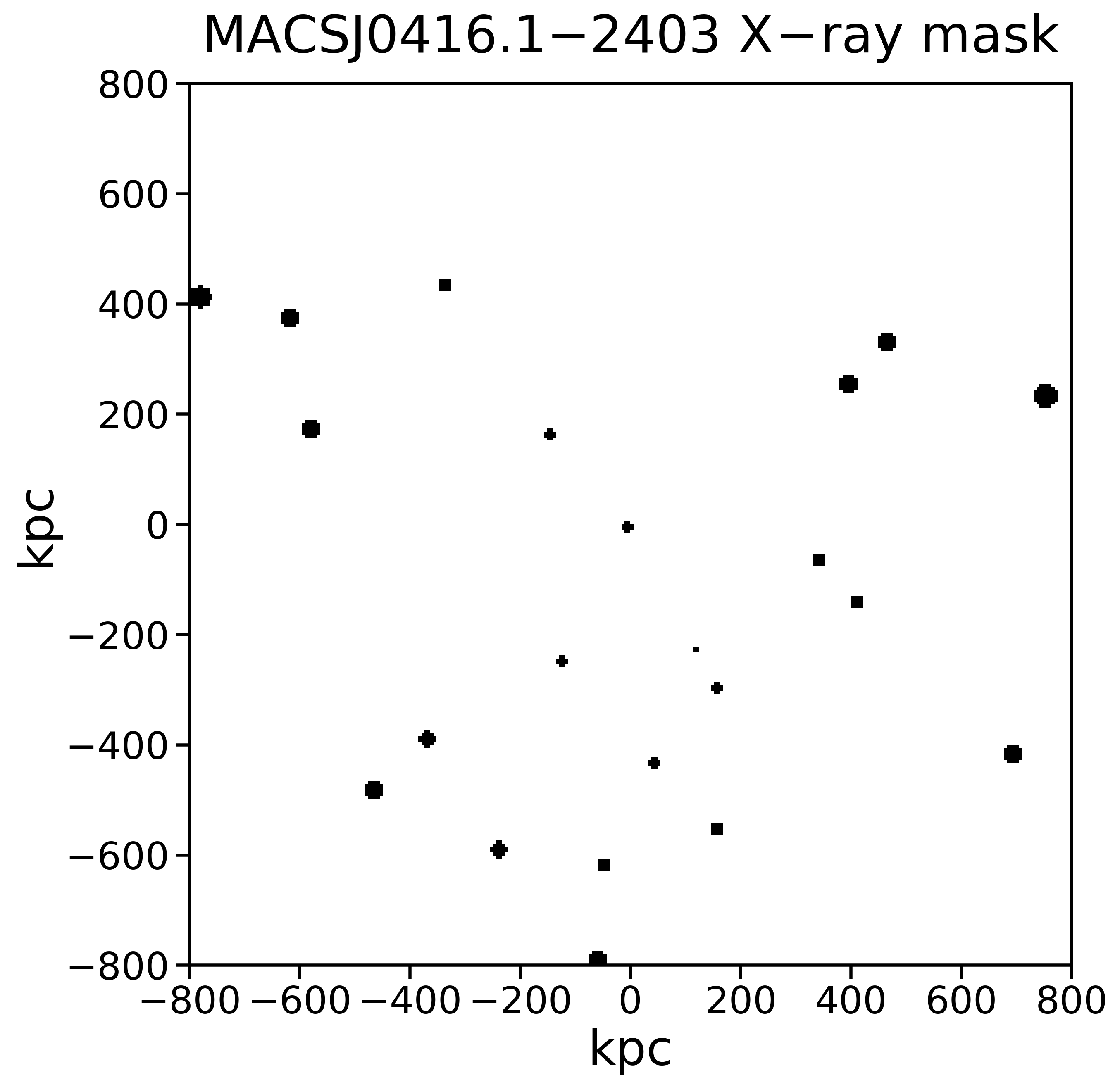} 
 \caption{Point-source mask for the galaxy cluster MACSJ0416.1$-$2403, corresponding to the system shown in Fig.~\ref{MACSJ0416.1-2403_coherence}.}
 \label{MACSJ0416_mask}
\end{figure}

These indicators are often used individually or in combination (e.g. \citealt{Rasia2013}, \citealt{Mantz2015}, \citealt{Lovisari2017}, \citealt{Cialone2018}, \citealt{Bartalucci2019}, \citealt{DeLuca21}, \citealt{Campitiello2022}) and were also recently employed in the first eROSITA all-sky survey to classify cluster dynamical states from X-ray morphology (\citealt{Sanders2025}). Complementary optical indicators include the offset between the brightest cluster galaxy (BCG) and the X-ray or SZ centroid, which traces the displacement between the stellar and gas mass peaks (\citealt{Jones1984}, \citealt{Lavoie2016}, \citealt{DeLuca21}, \citealt{Casas2024}), and the magnitude gap between the first and second brightest galaxies, often linked to the dynamical maturity of the system (\citealt{Lopes2018}). More recently, morphological analyses based on Zernike polynomials have been applied to SZ maps (\citealt{Capalbo2025}), providing a compact mathematical decomposition of cluster shapes and asymmetries that complements traditional morphology-based approaches.

 \begin{figure*}
\centering
      \subfigure[]{\includegraphics[width=0.33\textwidth]{good_figures/ABELL2261_kappa_zitrin_ltm_gauss_v2.png}}
      \subfigure[]{\includegraphics[width=0.33\textwidth]{good_figures/ABELL2261_xray_zitrin_ltm_gauss_v2.png}}
      \subfigure[]{\includegraphics[width=0.274\textwidth]{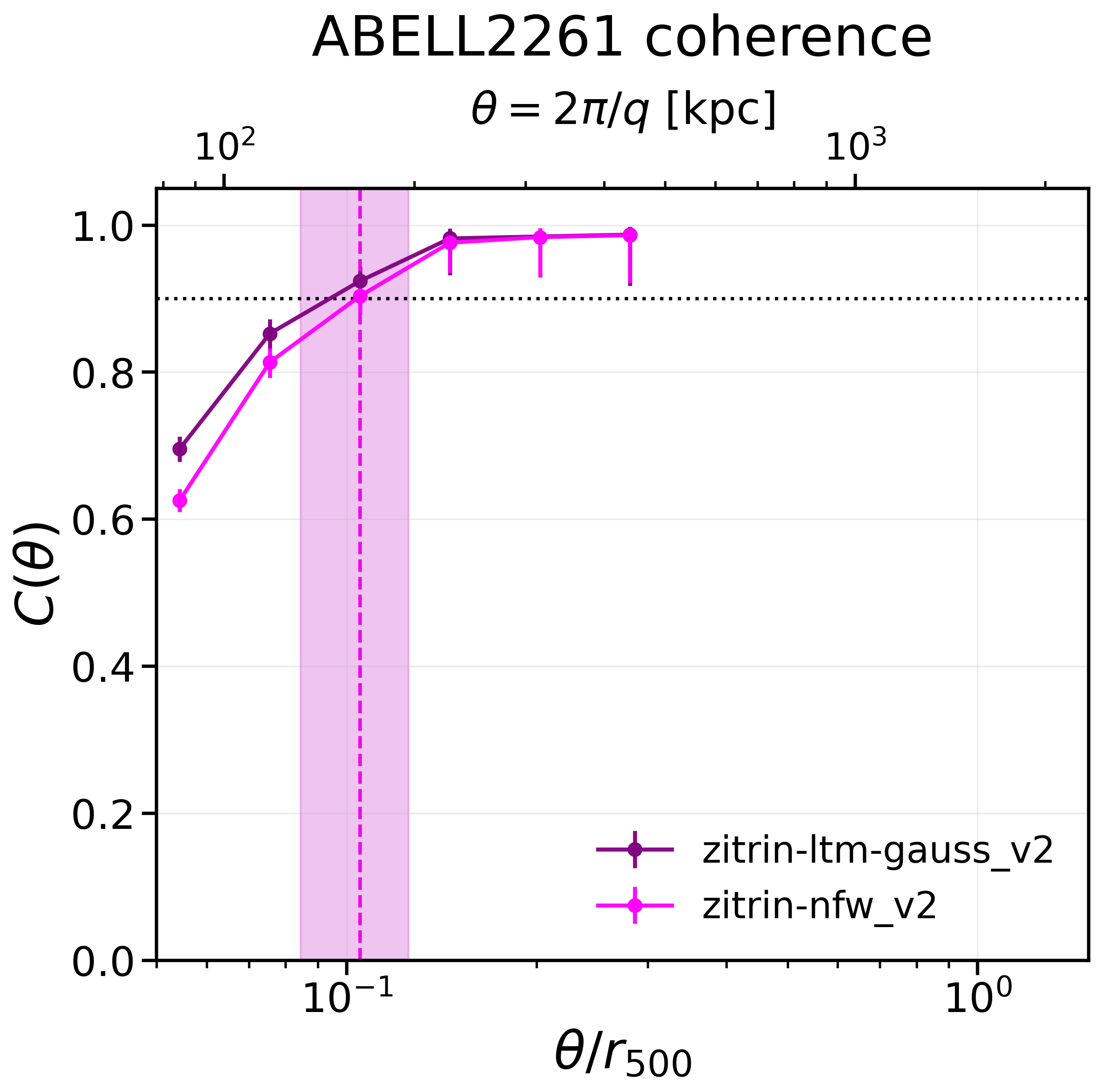}}     
\caption{Left panel: convergence map $\kappa(\boldsymbol{x})$ of the galaxy cluster ABELL2261 (CLASH) reconstructed with the \texttt{zitrin\_ltm\_gauss\_v2} model. Central panel: background-subtracted, exposure-corrected X-ray $(A+B)$ map of ABELL2261. Right panel: coherence measured for ABELL2261 using the \texttt{zitrin\_ltm\_gauss\_v2} and \texttt{zitrin\_nfw\_v2} lensing reconstructions, shown here as an example of a highly relaxed system.}
      
 \label{ABELL2261_coherence}
 \end{figure*}

In our previous works (\citealt{https://Cerini22}, \citealt{Cerini2025}), we presented a Fourier-based framework to characterize the dynamical state of galaxy clusters by jointly analyzing the spatial distributions of DM - reconstructed through gravitational lensing - and the ICM, traced by X-ray emission. The key innovation of this approach is the explicit inclusion of the DM component, which dominates the cluster potential and governs the evolution of the system. By coupling this component with the ICM within a Fourier-space analysis, the method provides a simultaneous, scale-resolved view of both distributions, offering a comprehensive description of cluster structure. In particular, the framework employs coherence analysis, defined through the combination of cross-power and auto-power spectra of the two fields, to quantify the degree of correlation - normalized to unity - between the DM and gas distributions as a function of the spatial scale. The power-spectrum formalism, widely used to describe spatial fluctuations in various astrophysical contexts, from cosmic backgrounds to ICM investigations (e.g. \citealt{Kashlinsky_2005, Churazov_2012, Cappelluti_2012, Cappelluti_2013, Helgason_2014, Cappelluti_2017, Eckert_2017, zhuravleva17, Li_2018, Kashlinsky_2018, Planck2024}), provides the statistical foundation of this approach. The key quantity introduced to measure the dynamical state is the \textsl{coherence length}, $\ell_{\mathrm{CR}}$, which identifies the scale above which the two components remain 90$\%$ coherent. Relaxed systems exhibit high coherence across all scales and, therefore, a lower \textsl{coherence length}, whereas disturbed clusters show a pronounced drop at small and intermediate scales and, consequently, a higher \textsl{coherence length}. This behavior reflects the fundamentally different physics governing the two components: the ICM interacts electromagnetically and responds to shocks, turbulence, and ram-pressure forces during mergers, while DM does not experience electromagnetic interactions and remains unaffected by these processes. As a result, the ICM can become spatially displaced from the underlying DM, lowering the coherence and revealing departures from equilibrium.

The effectiveness of this framework depends critically on the ability to map the DM distribution. Gravitational lensing has become a cornerstone technique for tracing DM in galaxy clusters and across a wide range of scales - from galactic halos to the cosmic web. Since the pioneering reconstructions of mass distributions through strong and weak lensing (\citealt{Walsh1979}; \citealt{Soucail1987}; \citealt{Tyson1990}; \citealt{Kaiser1993}), lensing has transformed our ability to study the dark Universe.
Today, this progress has culminated in an unprecedented era for gravitational-lensing science: high-resolution reconstructions from the Hubble Space Telescope (HST; e.g., \citealt{Postman_2012}, \citealt{Lotz_2017}, \citealt{Cerny_2018}) and the James Webb Space Telescope (JWST; e.g. \citealt{Finner_2023}, \citealt{Finner2025}, \citealt{Scognamiglio2026}) now reveal substructure within cluster-scale DM halos. In parallel, the Superpressure Balloon-borne Imaging Telescope (SuperBIT; \citealt{superbit_javier, Gill2024}) demonstrated near–space-quality imaging in angular resolution, pointing stability, and field of view during its 2023 superpressure balloon flight, proving the feasibility of lensing observations from a balloon platform at a fraction of the cost of traditional space missions (\citealt{Shaaban2022}, \citealt{McCleary2023}, \citealt{saha2026}). In addition, wide-field surveys such as Euclid (e.g. \citealt{Euclid2025}), the Vera C. Rubin Observatory (e.g. \citealt{Brough2020}), and the future Nancy Grace Roman Space Telescope (e.g. \citealt{Eifler2021}) are designed to map lensing signals over vast cosmological volumes, further extending the reach of this technique across the sky.

In this paper, we focus on the coherence-based framework for assessing the dynamical state of galaxy clusters. Our goal is to apply this method to a first representative sample of observed systems - the Hubble Space Telescope Frontier Fields (HSTFF\footnote{HSTFF lensing mass maps are publicly available at \href{https://doi.org/10.17909/T9KK5N}{DOI: 10.17909/T9KK5N}.}; \citealt{Lotz_2017}), the Cluster Lensing And Supernova Survey with Hubble (CLASH\footnote{CLASH lensing mass maps are publicly available at \href{https://doi.org/10.17909/T90W2B}{DOI: 10.17909/T90W2B}.}; \citealt{Postman_2012}), and the Reionization Lensing Cluster Survey (RELICS\footnote{RELICS lensing mass maps are publicly available at \href{https://doi.org/10.17909/T9SP45}{DOI: 10.17909/T9SP45}.}; \citealt{Cerny_2018}) - for which high-quality gravitational-lensing mass reconstructions are publicly available. These clusters span a broad redshift range, $0.19 < z < 0.97$, providing an ideal benchmark for testing the robustness of coherence-based dynamical-state indicators across different evolutionary stages. The full list of targets analyzed in this work is presented in Table~\ref{table:clusters_models}.

\begin{figure*}
\centering
\subfigure[]{\includegraphics[width=0.33\textwidth]{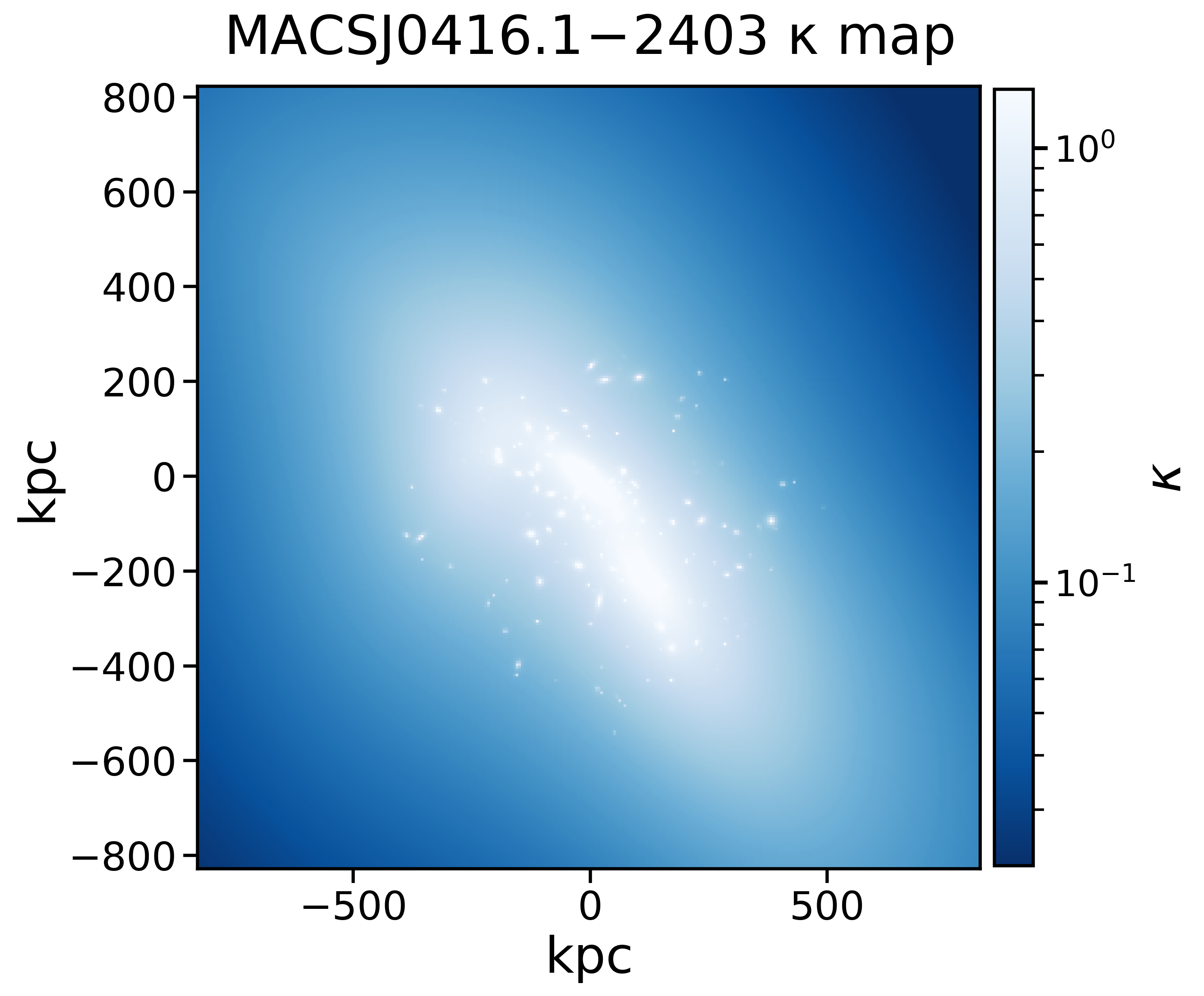}}
\subfigure[]{\includegraphics[width=0.33\textwidth]{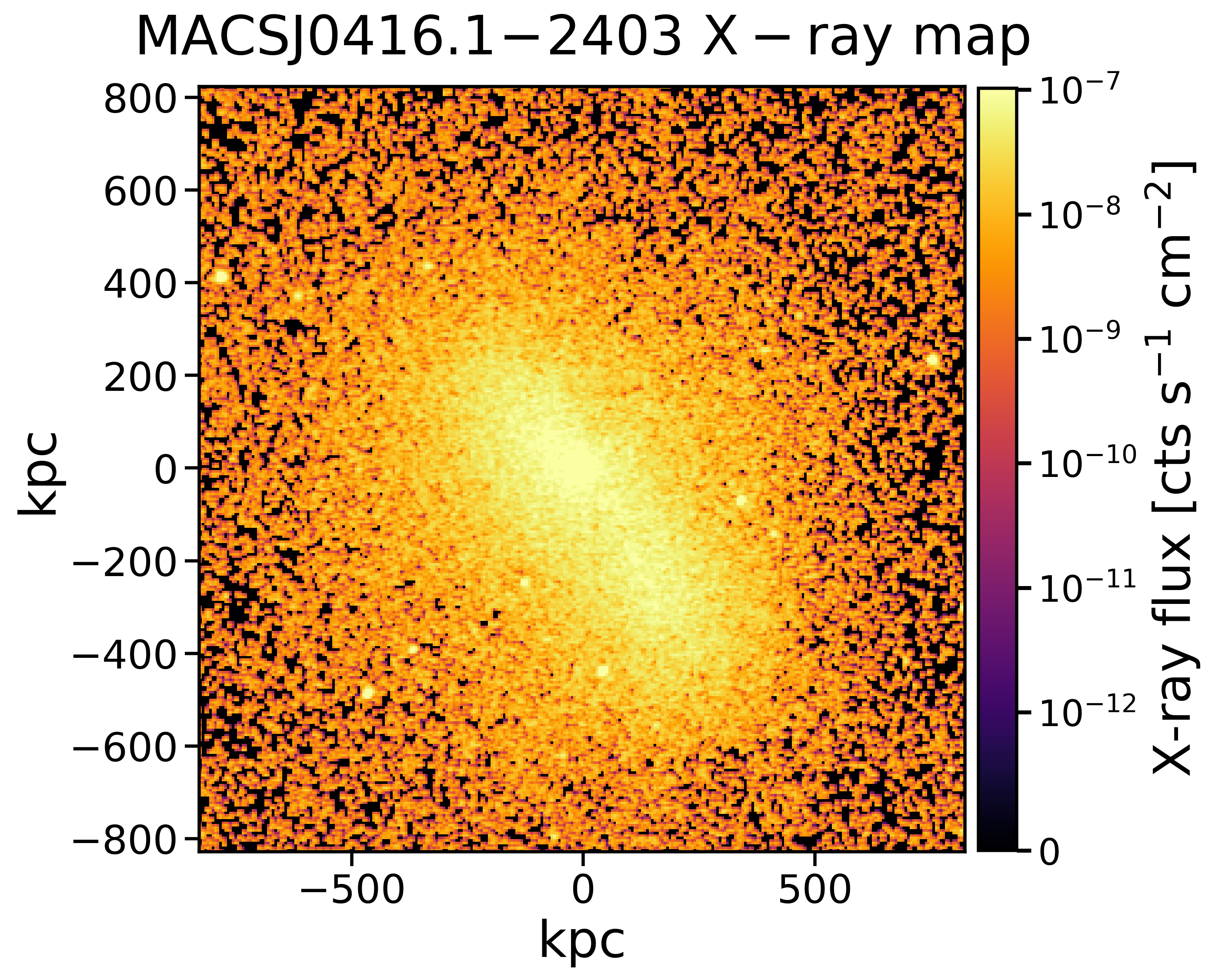}}
\subfigure[]{\includegraphics[width=0.27\textwidth]{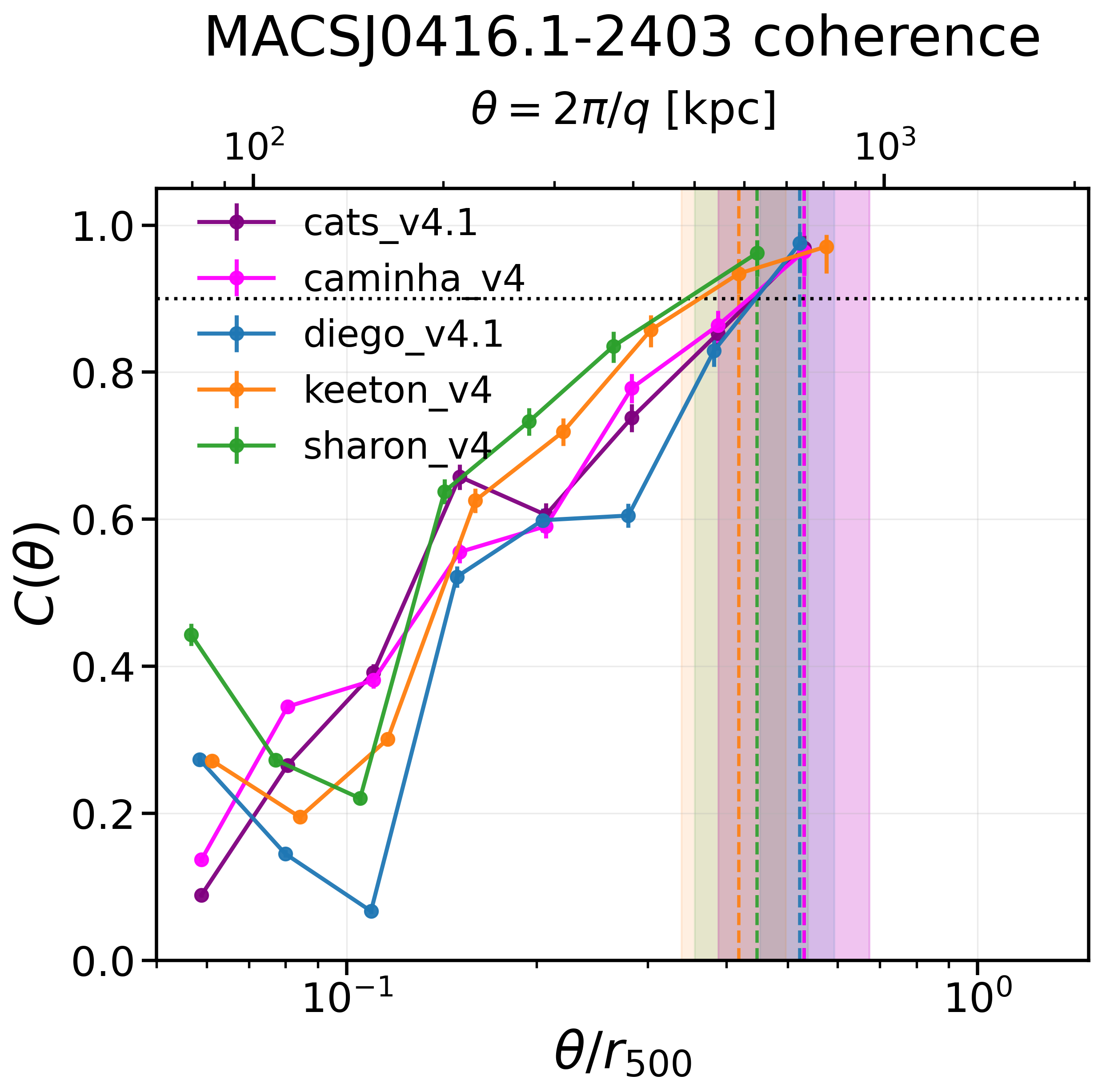}}
\caption{Left panel: convergence map $\kappa(\boldsymbol{x})$ of the galaxy cluster MACSJ0416.1$-$2403 (HSTFF), reconstructed with the \texttt{cats\_v4.1} model. Central panel: background-subtracted, exposure-corrected X-ray $(A+B)$ map of the same cluster. Right panel: coherence measured for MACSJ0416.1$-$2403 across the available lensing models, shown here as an illustrative example of an unrelaxed system.}
 \label{MACSJ0416.1-2403_coherence}
\end{figure*}

For the X-ray analysis, we use observations from the Chandra X-ray Observatory, whose on-axis half-power diameter (HPD) of $\sim 0\farcs5$ provides the highest angular resolution currently available in X-ray astronomy, ensuring optimal sensitivity to small-scale structure within the ICM. Over the redshift range of our sample, this angular resolution corresponds to physical scales of $\sim 1.6$ - $4.0$ kpc. This allows us to probe fluctuations in the X-ray surface brightness down to very small physical scales, extending the analysis to radii of only a few kpc. The high angular resolution of Chandra ensures that the scale at which the coherence drops below $C=0.9$---used to define the \textsl{coherence length}---is driven by the underlying physical structure of the cluster rather than by instrumental limitations. In particular, PSF smoothing does not bias the measurement of very small \textsl{coherence length}s in the most relaxed systems. All clusters are observed with the ACIS-I detector, whose wider field of view and larger effective area - compared to ACIS-S - make it particularly well suited for capturing the diffuse intracluster gas over cluster-wide scales. The analysis is performed in the broad 0.5-7 keV energy band, which optimizes the signal-to-noise ratio for hot ICM emission while minimizing contamination from soft Galactic foregrounds and high-energy particle backgrounds.

This study presents the first dynamical-state classification of observed clusters based on the coherence analysis and introduces the accompanying public code for coherence-based dynamical-state assessment, designed for broad applicability to future datasets; the code will be released upon acceptance of this paper. A comprehensive, quantitative comparison between this framework and other dynamical-state diagnostics - such as X-ray morphological estimators, centroid shifts, and optical indicators - will be presented in a follow-up work.

The paper is structured as follows: the data are described in Section~\ref{data}, the coherence analysis is described in Section~\ref{coherence}, results and conclusions are presented in Section
~\ref{results} and \ref{conclusions}, respectively. Throughout the paper, errors are quoted at 1$\sigma$ level unless otherwise specified. 

\begin{figure*}
\centering
\subfigure[]{\includegraphics[width=0.33\textwidth]{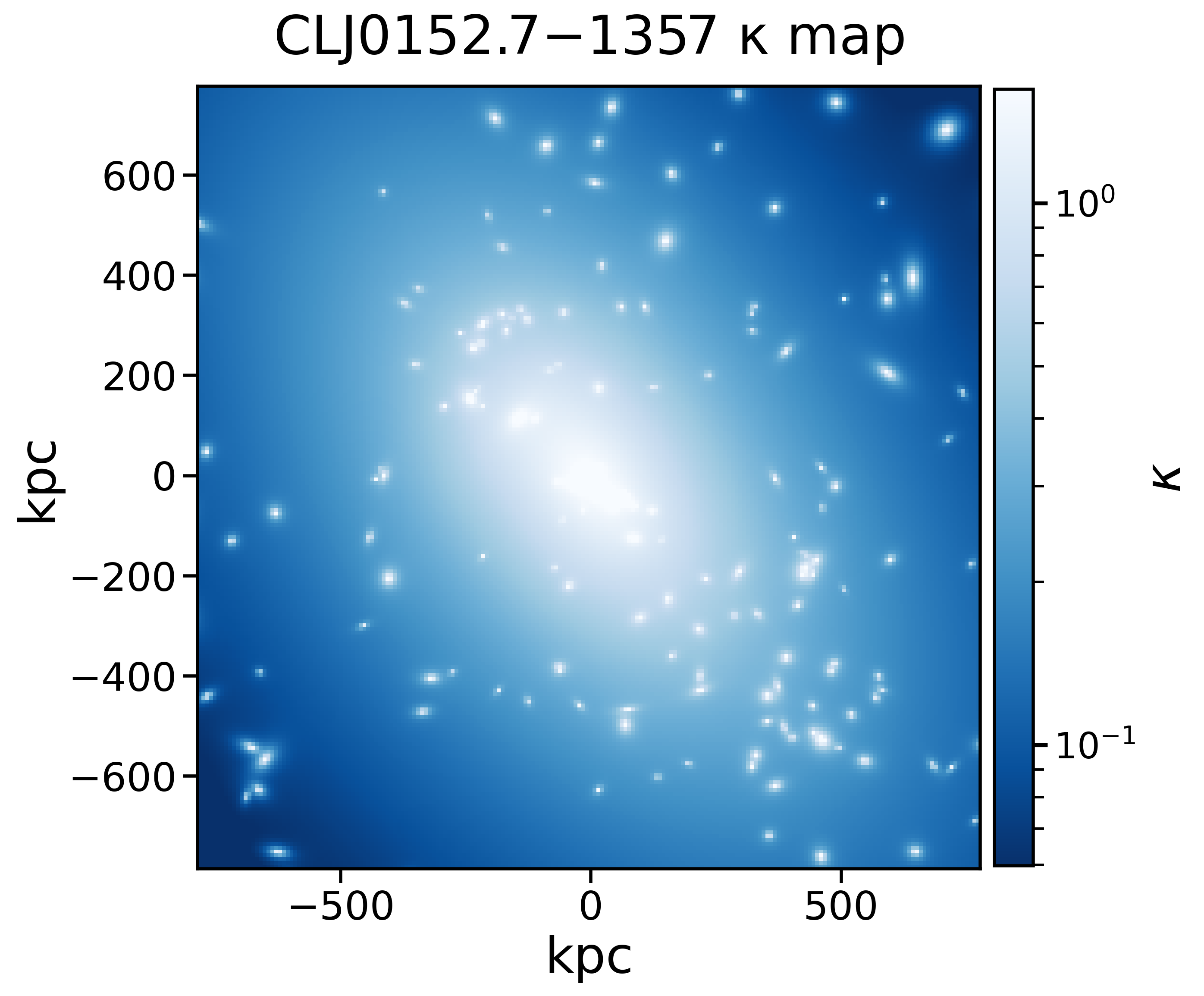}}
\subfigure[]{\includegraphics[width=0.33\textwidth]{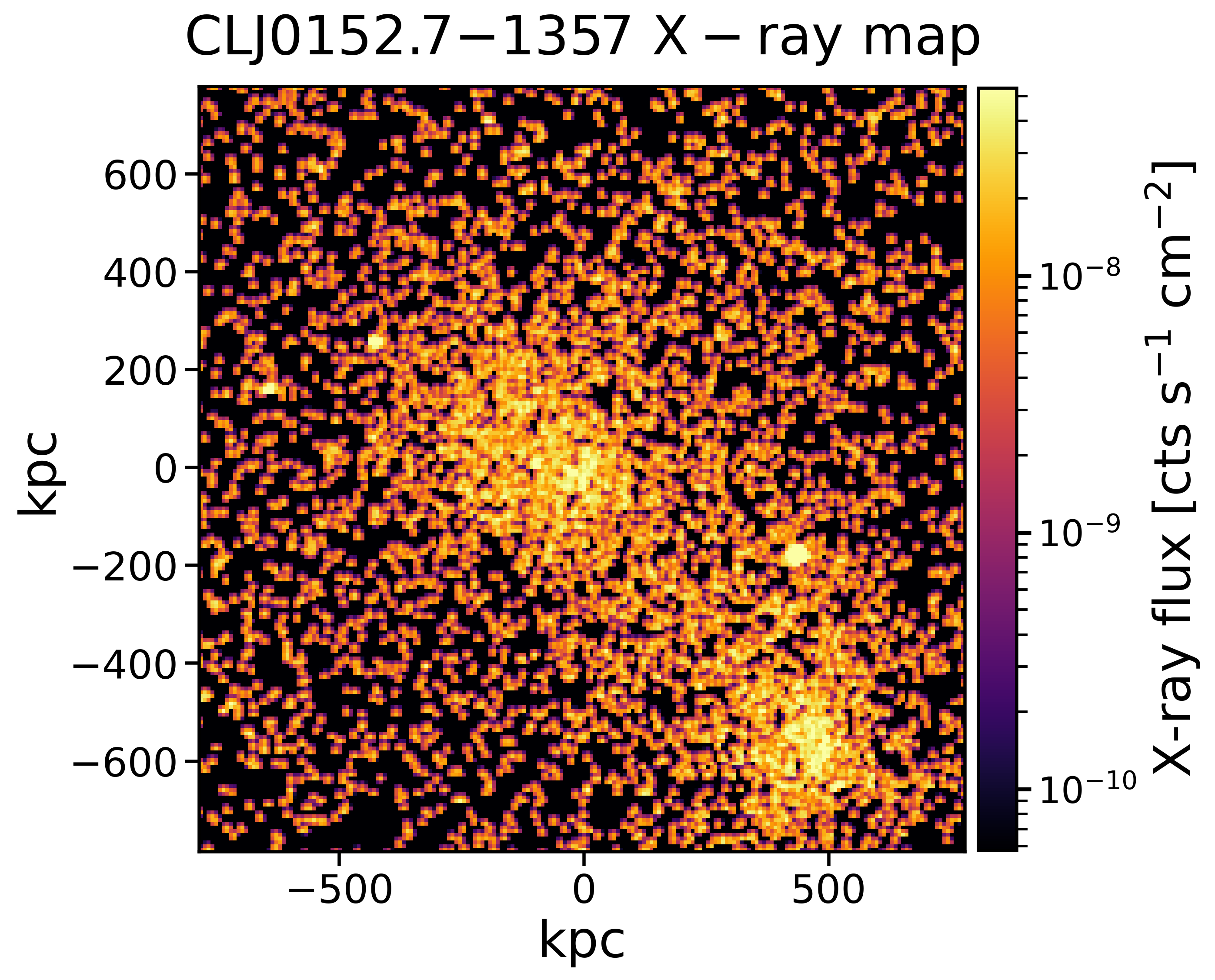}}
\subfigure[]{\includegraphics[width=0.27\textwidth]{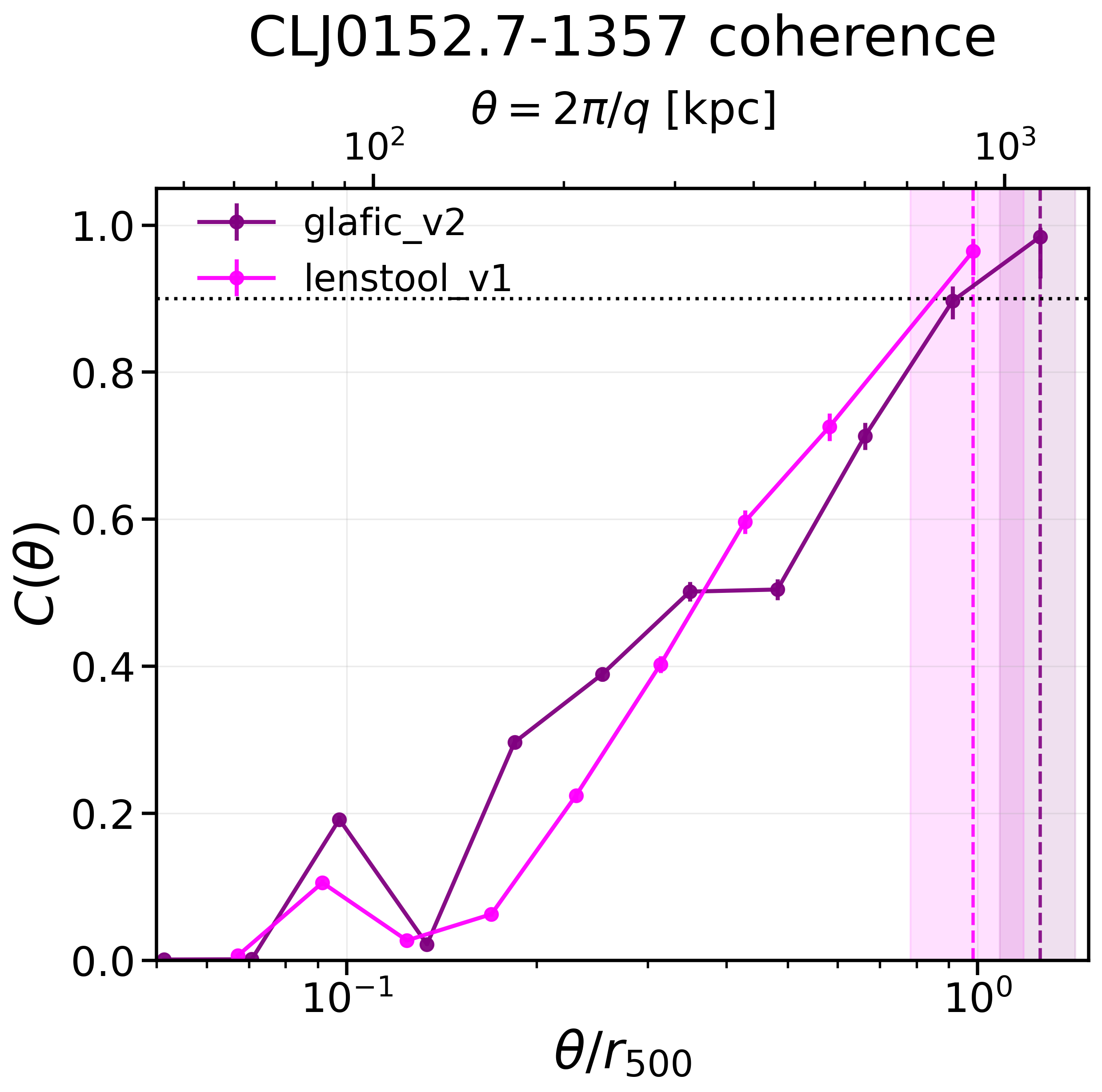}}
\caption{Left panel: convergence map $\kappa(\boldsymbol{x})$ of the galaxy cluster CLJ0152.7$-$1357 (RELICS), reconstructed with the \texttt{lenstool\_v1} model. Central panel: background-subtracted, exposure-corrected X-ray $(A+B)$ map of the same cluster. Right panel: coherence measured for CLJ0152.7$-$1357 across the available lensing models, shown here as an illustrative example of a highly disturbed (unrelaxed) system.}
\label{CLJ0152.7-1357_coherence}
\end{figure*}

\section{The Dataset} \label{data}
\subsection{Lensing mass reconstruction} \label{lensing_reconstruction}
The projected DM distribution in galaxy clusters can be reconstructed from the gravitational lensing distortions they imprint on background galaxies. The deep gravitational potential of a cluster curves the surrounding space-time, causing light from more distant sources to be deflected as it travels toward the observer. Because gravitational lensing is a purely geometric effect arising from the curvature of space-time, it is achromatic and depends only on the mass distribution along the line of sight.

When the observer, lens, and background source are nearly aligned and the cluster has a sufficiently deep and centrally concentrated gravitational potential, the system enters the strong-lensing regime. In this case, the lens mapping becomes highly non-linear: background galaxies can be multiply imaged, strongly magnified, and stretched into giant arcs or Einstein rings. Mass reconstructions in this regime typically rely on solving the full lens equation using parametric or free-form models constrained by the positions, shapes, and redshifts of multiple images (for a comprehensive review of strong-lensing techniques in galaxy clusters, see \citealt{KneibNatarajan_2011}).

For more modest alignments, the induced distortions are much smaller and coherent only at the statistical level, corresponding to the weak-lensing regime. Here, the mapping between source and image can be linearized, and the primary observable is the shear field, inferred statistically from the averaged ellipticities of large samples of background galaxies after correcting for point-spread-function effects and shape-measurement biases (for a comprehensive review of weak-lensing techniques, see \citealt{Kaiser1993}, \citealt{Bartelmann:1999}).

Strong and weak lensing therefore provide complementary information: strong lensing tightly constrains the mass distribution in the dense central regions of clusters, while weak lensing traces the mass over larger scales and at lower surface densities, enabling a joint reconstruction of the full projected mass distribution, dominated by DM. Strong-lensing reconstructions can achieve percent-level precision (∼1--2$\%$) in the inferred convergence maps; however, they rely on specific modeling assumptions, and the impact of these assumptions will be discussed in this work. In contrast, weak-lensing convergence maps do not depend on any model of the cluster potential and are therefore less sensitive to modeling systematics. A detailed analysis based on weak-lensing maps will be presented in a follow-up paper.

In this work, for all clusters in our sample, we use the publicly available convergence maps $\kappa(\boldsymbol{x)}$, defined as

\begin{equation}
    \kappa(\boldsymbol{x}) = 
    \frac{\Sigma(\boldsymbol{x})}{\Sigma_{\mathrm{crit}}},
    \label{kappa_def}
\end{equation}

where $\boldsymbol{x}$ denotes the angular position on the sky, $\Sigma(\boldsymbol{x})$ is the projected surface mass density of the cluster and $\Sigma_{crit}$ is the critical surface density for gravitational lensing,

\begin{equation}
    \Sigma_{\mathrm{crit}} = 
    \frac{c^{2}}{4\pi G}\,
    \frac{D_{\mathrm{s}}}{D_{\mathrm{l}} D_{\mathrm{ls}}},
    \label{sigma_crit}
\end{equation}

with $D_{\mathrm{l}}$, $D_{\mathrm{s}}$ and $D_{\mathrm{ls}}$
the angular-diameter distances to the lens, to the source, and from the lens to the source. The convergence $\kappa$ is therefore a dimensionless measure of the projected mass density, representing the fluctuations of $\Sigma$ scaled by the lensing efficiency. Regions where $\Sigma>\Sigma_{crit}$
can produce multiple images, while regions with $\Sigma<\Sigma_{crit}$ generate only weak distortions.

Below, we outline the relevant observational properties of the three cluster samples employed in this work; a summary of the full sample is provided in Table~\ref{table:clusters_models}. In particular, we restrict our analysis to models with comparable angular resolution ($\sim$1 arcsec) and map extent, ensuring a fair comparison across different reconstructions. The full list of models adopted in this work is reported in Table~\ref{table:clusters_models}.

\begin{figure*}
\centering
\subfigure[]{\includegraphics[width=0.32\textwidth]{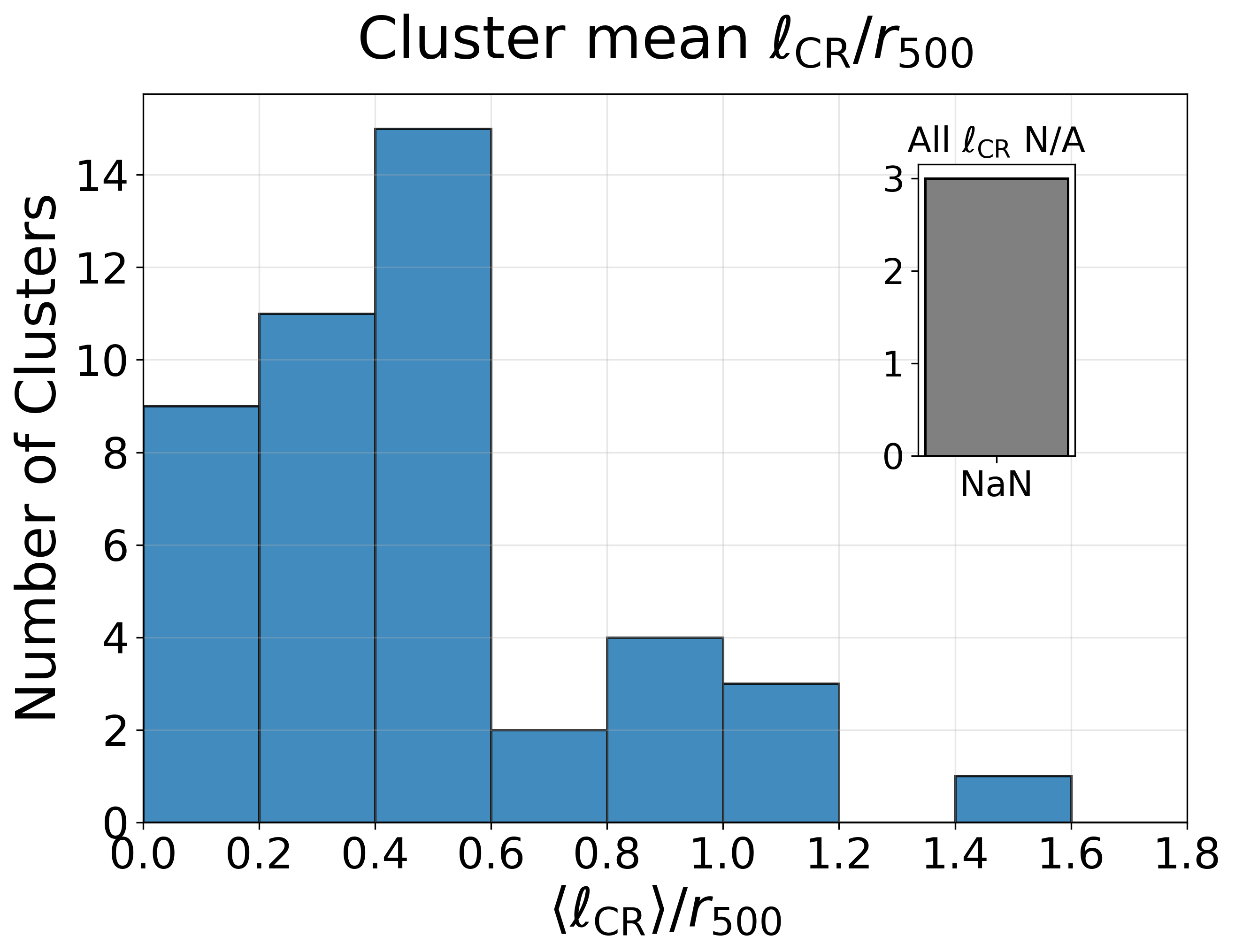}}
\subfigure[]{\includegraphics[width=0.32\textwidth]{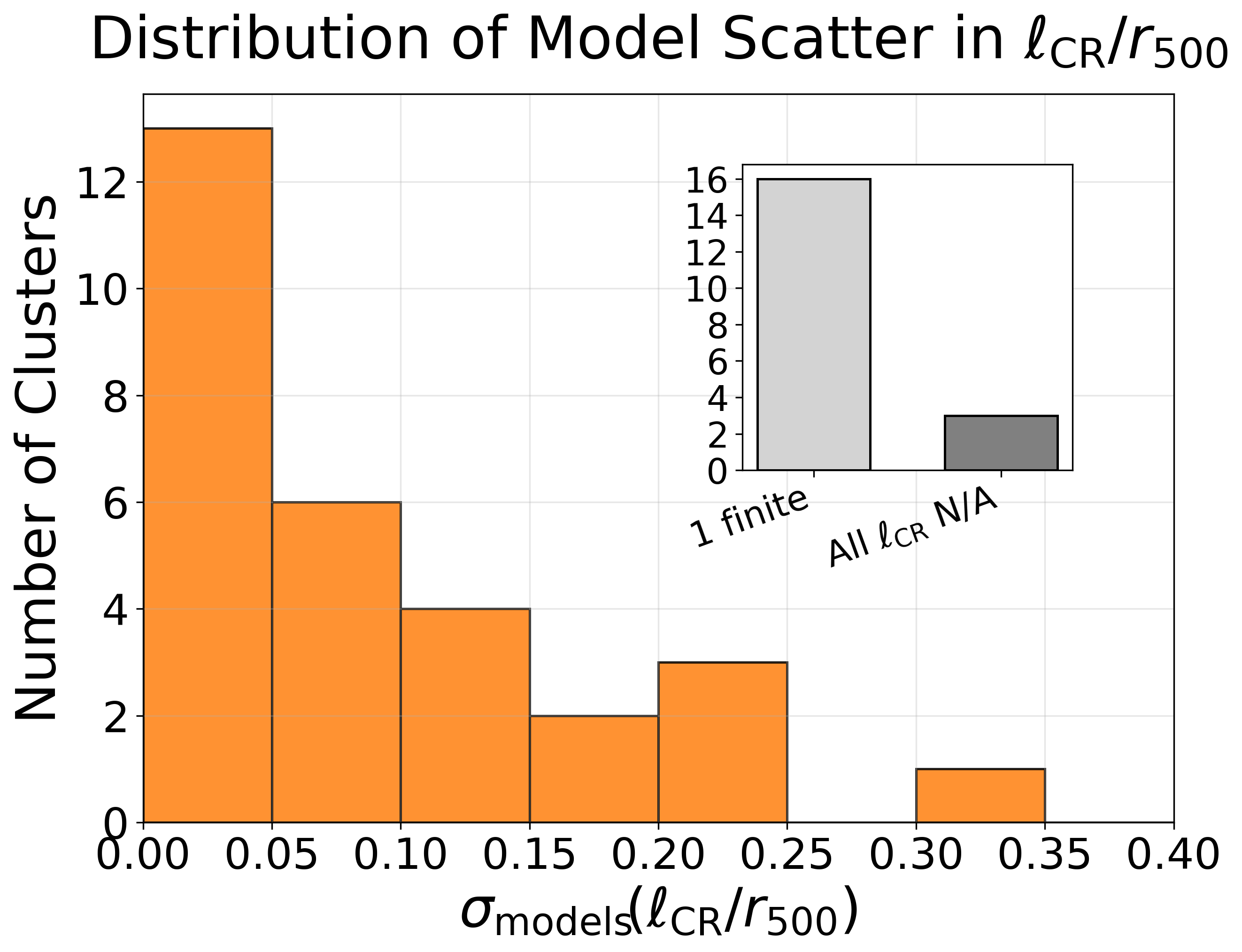}}
\subfigure[]{\includegraphics[width=0.32\textwidth]{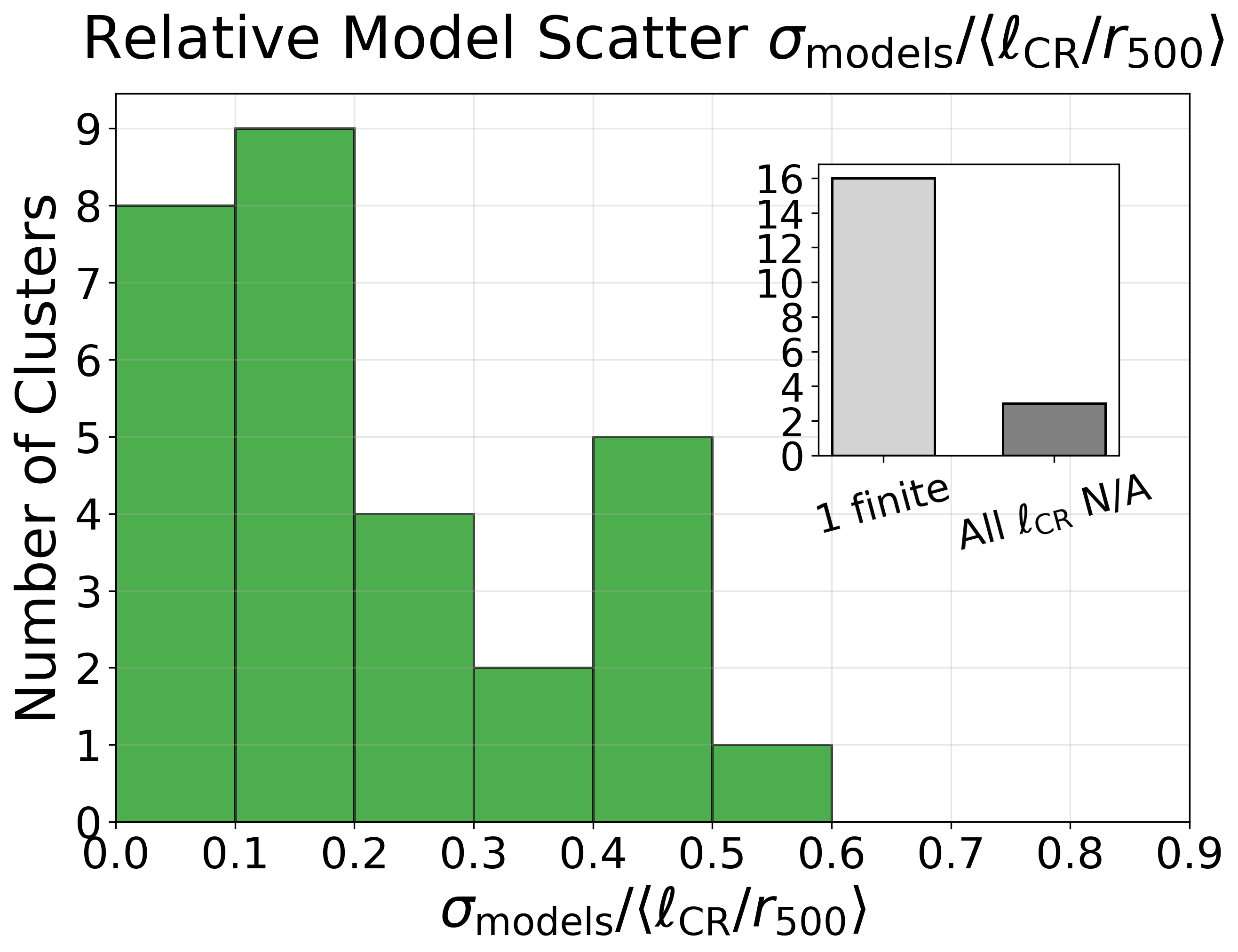}}
\caption{Distribution of the normalized \textsl{coherence length} across the full cluster sample. 
\textit{Left:} histogram of the mean $\langle \ell_{\rm CR}/r_{500} \rangle$ computed for each cluster across all available lensing models. 
\textit{Middle:} distribution of the model-to-model scatter, quantified by $\sigma_{\mathrm{models}}$. 
\textit{Right:} distribution of the relative scatter, expressed as $\sigma_{\mathrm{models}} / \langle \ell_{\rm CR}/r_{500} \rangle$. 
The insets highlight clusters for which the \textsl{coherence length} cannot be robustly determined across all models: dark grey bars indicate systems for which no $\ell_{\rm CR}$ is defined (i.e.\ the coherence does not reach $C=0.9$ for any model), while light grey bars correspond to clusters with a finite $\ell_{\rm CR}$ measured for only one model.
}
\label{overall_distr}
\end{figure*}

\subsection{HST lensing maps} \label{lens_data}

The HSTFF program (\citealt{Lotz_2017}) targeted six massive galaxy clusters selected for their exceptional lensing strength. In this work, we consider only the subset of publicly available HSTFF lens models listed in Table~\ref{table:clusters_models}, namely the CATS, Diego, Keeton, Sharon, Williams, Zitrin-LTM-Gauss, Zitrin-LTM, and Zitrin-NFW reconstructions. The CATS and Sharon models are parametric reconstructions generated with the \textsc{Lenstool} software (\citealt{Jullo_2007}, \citealt{Jullo_2009}), in which the cluster mass distribution is described as a combination of large-scale halos and galaxy-scale perturbers optimized to reproduce the observed lensing constraints. The Diego models are based on the free-form \textsc{WSLAP+} method (\citealt{2007MNRAS.375..958D}), which reconstructs the mass distribution on a flexible grid without imposing global analytic profiles. The Williams models are obtained with the free-form \textsc{GRALE} algorithm (\citealt{2006MNRAS.367.1209L}, \citealt{Sebesta_2016}, \citealt{2017MNRAS.465.1030P}), while the Keeton models are parametric reconstructions produced with \textsc{LENSMODEL}. The Zitrin models adopt either a Light-Traces-Mass (LTM) approach (\citealt{Broadhurst_2005}, \citealt{Zitrin_2009}) or an alternative analytical NFW-based parameterization (\citealt{2013ApJ...762L..30Z}). All HSTFF reconstructions are based on deep ACS and WFC3/IR imaging obtained across multiple filters.

The CLASH sample (\citealt{Postman_2012}) consists of  galaxy clusters observed with a 16-band \textit{HST} program designed to enable precise photometric redshifts and high-fidelity lensing reconstructions. The sample was primarily selected based on X-ray morphological regularity, and is therefore dominated by dynamically relaxed, cool-core systems, with a smaller subset of clusters chosen for their strong-lensing efficiency. For the present analysis, we use the CLASH models listed in Table~\ref{table:clusters_models}, namely the Zitrin-LTM-Gauss, Zitrin-LTM, and Zitrin-NFW reconstructions. These models are based on the lensing reconstruction framework developed by Zitrin et al. (e.g.~\citealt{Zitrin_2011}) and released as part of the CLASH lens-modeling effort. In this framework, one family of models adopts an LTM-based description of the mass distribution, while the other uses an analytical elliptical NFW parameterization for the dark-matter component. These complementary approaches were released to assess systematic uncertainties associated with lens-model assumptions.

Finally, the RELICS sample (\citealt{Cerny_2018}, \citealt{Coe_2019}, \citealt{Salmon_2020}) includes massive galaxy clusters spanning a wider range of redshifts and dynamical states. In this work, we restrict the analysis to the RELICS lens-model families listed in Table~\ref{table:clusters_models}, namely GLAFIC, Lenstool, Zitrin-LTM-Gauss, and Zitrin-NFW. These publicly available reconstructions are derived primarily from strong-lensing constraints identified in the RELICS \textit{HST} imaging. The GLAFIC models are parametric reconstructions generated with the \textsc{GLAFIC} code (\citealt{2010PASJ...62.1017O}, \citealt{Ishigaki_2015}, \citealt{Kawamata_2018}), while the Lenstool models are produced with the \textsc{Lenstool} software (\citealt{Jullo_2007}, \citealt{Jullo_2009}). The Zitrin models follow either the LTM approach or an NFW-based parameterization, providing complementary descriptions of the cluster mass distribution. We do not include additional model families that are not uniformly available across the RELICS sample or that are not listed in Table~\ref{table:clusters_models}.

For all clusters and lensing models, associated uncertainty maps for the convergence ($\kappa$) are also provided. These correspond to $1\sigma$ error maps derived from the posterior distribution of the lens models, estimated through Monte Carlo sampling of the Markov Chain Monte Carlo (MCMC) chains used in the reconstruction. In practice, they are obtained from ensembles of realizations of the lens model, capturing the statistical uncertainties driven by the lensing constraints. Both the convergence maps and their associated uncertainties are used in the computation of the power spectra and coherence, as described in Section~\ref{coherence}. An example of a convergence map and its associated uncertainty map for ABELL2261, reconstructed with the \texttt{zitrin\_ltm\_gauss\_v2} model, is shown in Fig.~\ref{kappa_example}. As previously anticipated, reconstruction techniques that include strong-lensing constraints can achieve uncertainties at the $\sim$1--2\% level, resulting in a noise power that is approximately $10^{-4}$ lower than that of the signal.

In \citealt{https://Cerini22}, we found that, for the HSTFF cluster Abell~2744 and the CLASH cluster Abell~383, the inferred \textsl{coherence lengths} were not strongly affected by the choice of lensing model. In this work, we revisit this aspect in a systematic way by comparing multiple lensing reconstructions across all clusters in our sample.

\begin{figure*}
\centering
\subfigure[]{\includegraphics[width=0.33\textwidth]{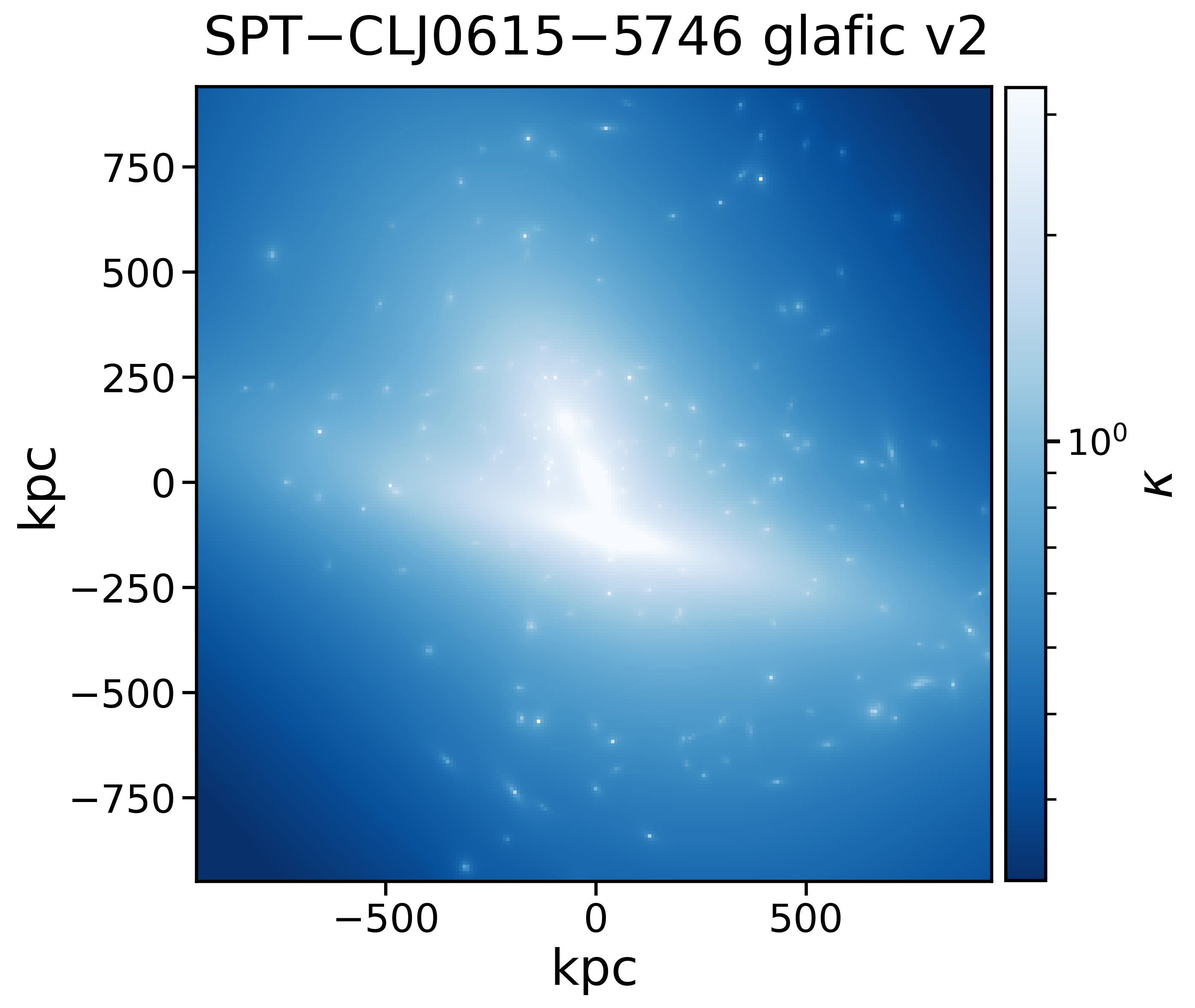}}
\subfigure[]{\includegraphics[width=0.33\textwidth]{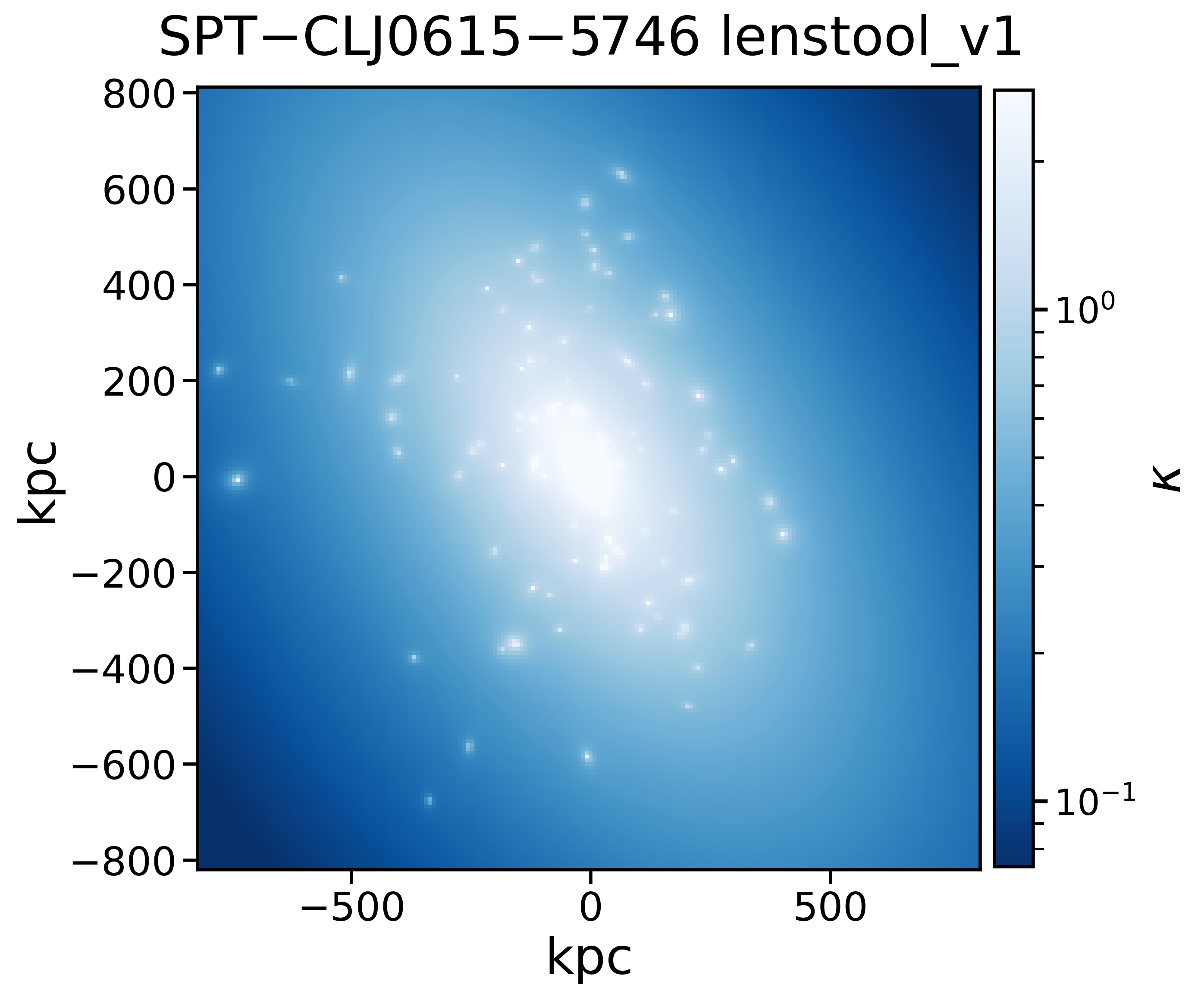}}
\subfigure[]{\includegraphics[width=0.275\textwidth]{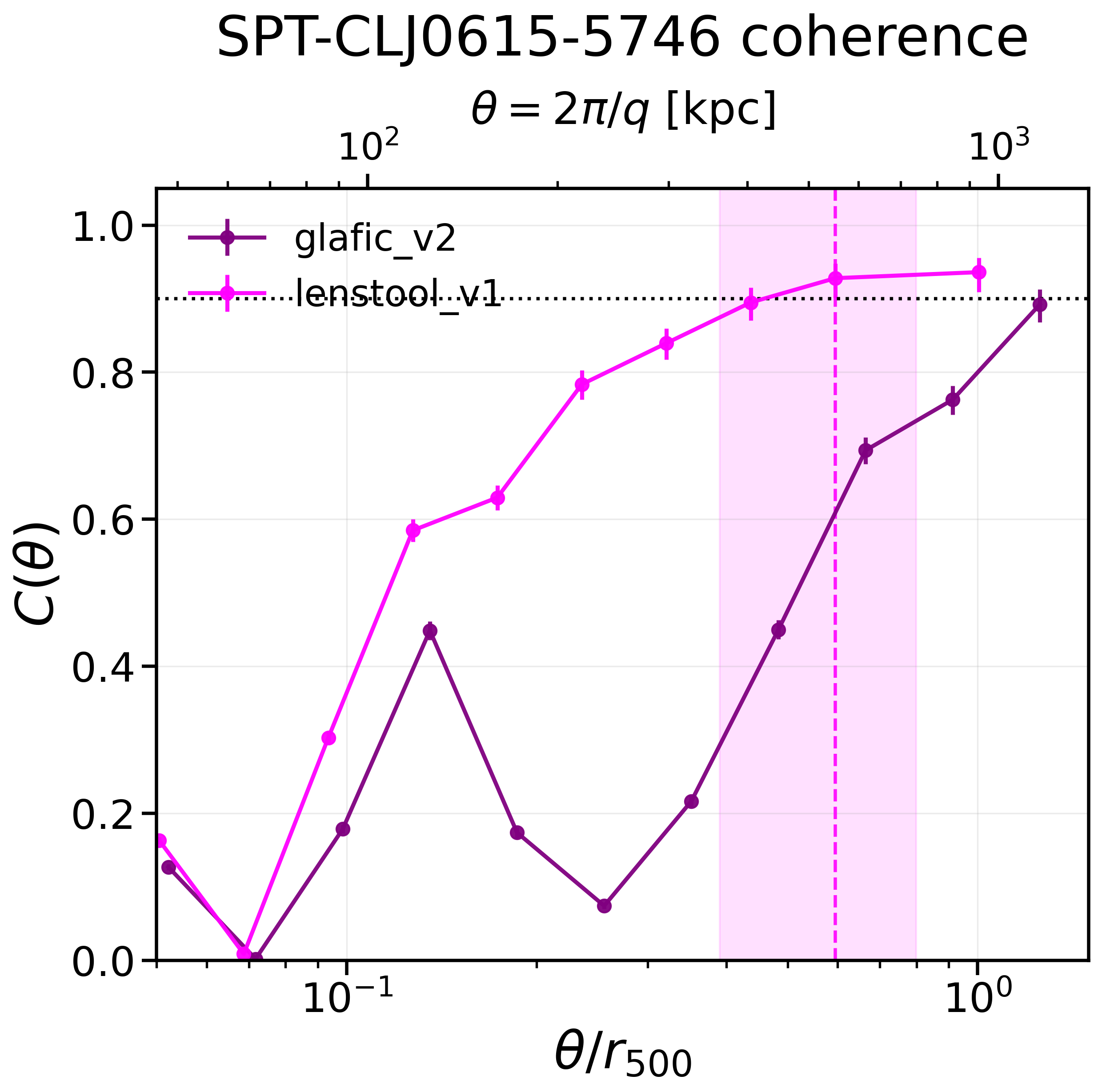}}
\caption{Example of significant model-dependent variation in the coherence signal for the cluster SPT-CLJ0615$-$5746. The left and central panels show the $\kappa$ maps obtained from two different lensing reconstructions, \texttt{glafic\_v2} and \texttt{lenstool\_v1}, respectively. The right panel presents the corresponding coherence measurements, highlighting the differences in the inferred signal and \textsl{coherence length} between the two models.}
\label{SPT_example}
\end{figure*}

%\noindent\mbox{}\par
\subsection{X-ray data} \label{xray}

For the X-ray images, we used Chandra ACIS-I observations in the 0.5--7.0 keV energy band. For each cluster, all available observations (ObsIDs) were reprocessed starting from the level-1 event files using standard procedures with the Chandra Interactive Analysis of Observations (CIAO; version 4.17) software and the latest calibration database (CALDB; version 4.11). In particular, the level-1 event files are reprocessed using \texttt{chandra\_repro} to apply the most up-to-date calibration, remove bad pixels and cosmic-ray afterglows, and filter events by grade and status to retain only scientifically valid detections. Periods of enhanced particle background (“flares’’) are identified through light-curve inspection and excluded, ensuring a temporally stable background. 

The background contribution, including both the instrumental particle background and the diffuse sky background, is modeled using blank-sky datasets. These are archival background event files constructed from observations of regions without bright target emission, and therefore provide an empirical estimate of the non-source background recorded by ACIS, including both particle-induced events and an average sky background component. The blank-sky files are reprojected and processed consistently with the science observations, and are renormalized using the count-rate ratio in the 9.5--12 keV band, where the Chandra effective area is negligible and the detected events are dominated by the particle background. This renormalization therefore matches the instrumental background level in each observation, accounting for its temporal variability, while the sky background component is left unchanged. This approximation is appropriate for the present analysis also because, unlike traditional fluctuation studies of the cosmic X-ray background (e.g.\citealt{Kashlinsky_2018}), we do not subtract the mean sky emission and analyze residual fluctuations. Similarly, in contrast to studies of surface-brightness fluctuations in galaxy clusters that remove a smooth underlying model (e.g.\ a $\beta$-model; \citealt{Churazov_2012}), we retain the full emission without subtracting a large-scale profile, as described in the next section. In this regime, and in particular within $r_{500}$ where our analysis is performed, the signal is strongly dominated by the ICM emission. X-ray studies of cluster outskirts consistently show that the outer regions become background dominated as one approaches the virial radius and beyond, whereas the ICM can still be traced robustly out to $\sim r_{500}$ and beyond in deep observations (e.g. \citealt{Vikhlinin2006}, \citealt{Ettori2010}, \citealt{Ghirardini2019}).  The dominant large-scale components of the diffuse sky background --- namely solar-wind charge exchange (SWCX), the local hot bubble (LHB), and galactic halo emission --- are expected to vary primarily on angular scales much larger than those probed in this work. As a result, across the spatial scales sampled here they can be approximated as a nearly uniform contribution, effectively acting as a constant offset. Their impact therefore enters primarily through the Poisson noise budget, as quantified below in this subsection, rather than introducing scale-dependent structure in the power spectrum. While SWCX can exhibit temporal variability between observations, it contributes predominantly at energies $\lesssim 1$ keV and, over the angular scales considered here, likewise behaves as an approximately uniform component across the field of view. At higher energies, the sky background is dominated by unresolved extragalactic emission, arising mainly from faint active galactic nuclei (AGN), galaxies, and diffuse intergalactic gas. This component can in principle show structure on the angular scales probed here. Indeed, in the unresolved 0.5--2 keV CXB power spectrum, \citealt{Cappelluti_2012} find excess power above shot noise on scales $>30''$, with the clustered extragalactic contribution becoming important on scales $\gtrsim 100''$ and remaining relevant out to $\sim 1000''$. Over the redshift range of our sample, $z\simeq0.19$--$0.97$, this corresponds to approximately $\sim 300$--$8000$ kpc. However, because we analyze the full X-ray surface-brightness distribution of galaxy clusters, where the ICM emission is strongly dominant within $r_{500}$, the contribution of these extragalactic fluctuations is subdominant and does not significantly affect the measured power spectrum or coherence. The use of blank-sky datasets therefore provides an adequate representation of the average sky background for the purposes of this work. The renormalization based on the 9.5--12 keV band exploits the empirical stability of the quiescent particle-background spectral shape (e.g. \citealt{Hickox_2006}).

\begin{figure}[ht]
 \centering
 \includegraphics[width=\columnwidth]{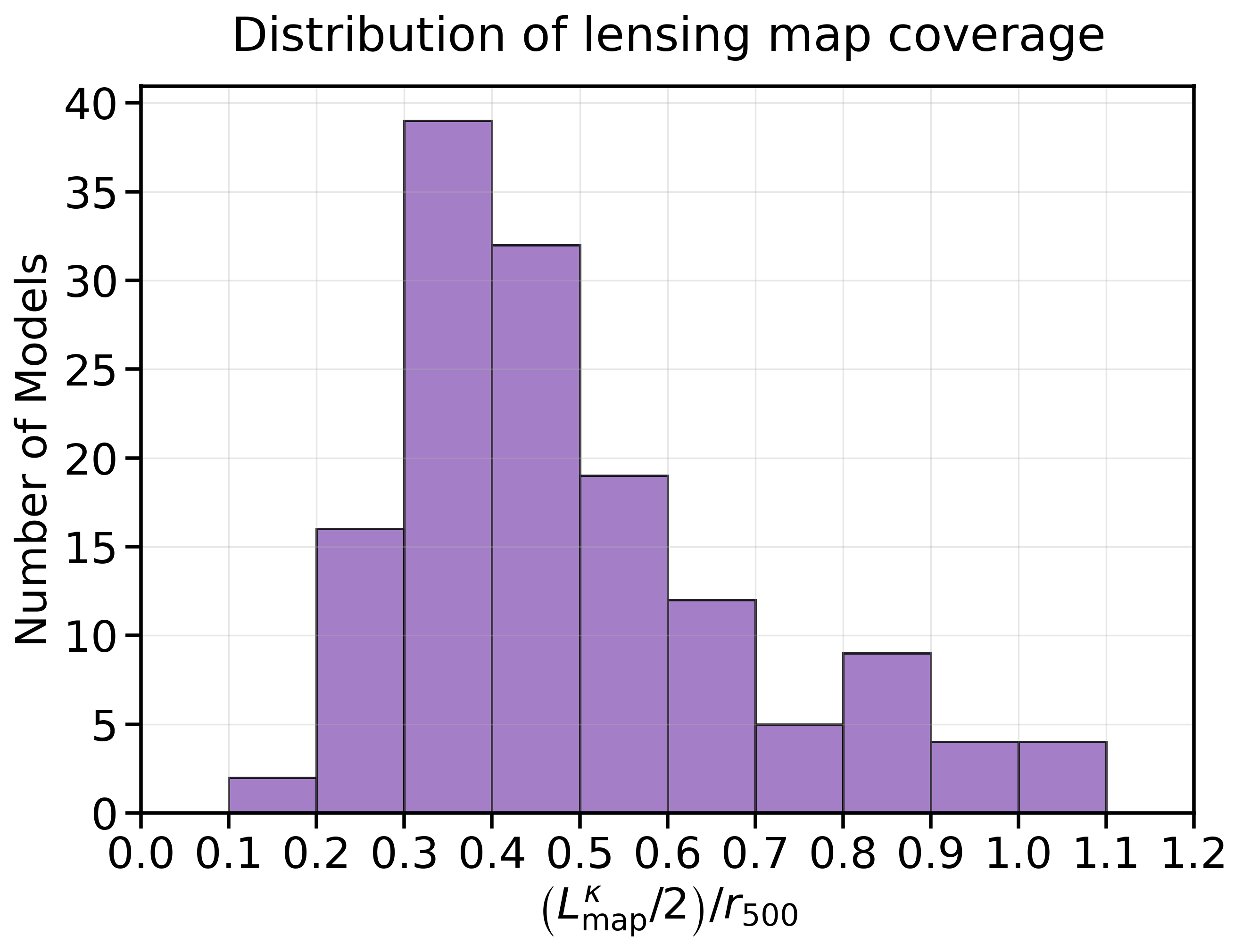} 
 \caption{Distribution of the lensing map extent for all models analyzed in this work. The quantity $L_{\rm map}^{\kappa}$ denotes the side length of the convergence map, so that $L_{\rm map}^{\kappa}/2$ corresponds to half of the map extent, measured from the map center. This value is normalized by $r_{500}$ to show the spatial coverage of each lensing reconstruction relative to the cluster size.}
 \label{size_map}
\end{figure}
      
After cleaning, counts images and exposure maps are generated in the 0.5--7 keV energy band using \texttt{fluximage}. Exposure maps account for variations in telescope effective area, detector quantum efficiency, chip gaps, and vignetting, enabling the conversion of raw counts to flux-normalized images suitable for spatial fluctuation analyses. In constructing the exposure maps, an appropriate spectral weighting is adopted to account for the energy dependence of the instrument response. Individual observations are then reprojected onto a common astrometric frame, and all images are produced on a common grid and pixel scale, ensuring consistency across ObsIDs. Exposure-corrected images are subsequently combined into a final mosaic. The reprocessed event files incorporate the telescope dithering pattern, preventing spurious structures in the subsequent Fourier analysis. The total exposure time varies across clusters. However, the use of exposure maps ensures that all images are consistently flux-calibrated and directly comparable across the sample.

To construct the maps used in the power-spectrum and coherence analysis, each Chandra event list was also split by photon arrival time into two interleaved subsets, hereafter $A$ and $B$. Each subset was converted into a counts image and divided by its corresponding exposure map. By construction, the two images have identical exposure, contain statistically equivalent realizations of the underlying astrophysical emission, and each carries half of the total photons. Thus, while 
$A$ and $B$ encode the same sky signal, their Poisson noise is independent.
From these, we formed the standard “sum’’ and “difference’’ maps, defined as 
$\tfrac{1}{2}(A+B)$ and $\tfrac{1}{2}(A-B)$, respectively. The “sum’’ map contains the astrophysical signal together with its associated noise, whereas in the “difference’’ map the true sky signal cancels out and only the noise component remains. This provides a clean way to characterize the intrinsic noise properties in the data. The $A$ - $B$ method is a well-established technique in X-ray surface-brightness fluctuation studies and has been widely used in previous analyses (e.g. \citealt{Kashlinsky2005}; \citealt{Cappelluti_2013}; \citealt{Cappelluti_2017}; \citealt{Li_2018}). An example of the power spectra of the $\tfrac{1}{2}(A+B)$ and $\tfrac{1}{2}(A-B)$ X-ray maps for ABELL2261 is shown in Fig.~\ref{xray_example} (the corresponding convergence map and the associated uncertainty map of the convergence field are presented in Fig.~\ref{kappa_example}). As described in the next section, Poisson noise dominates the power on small scales; its contribution can be removed by subtracting the noise power spectrum from the total, yielding the cleaned signal.

Finally, point sources — which introduce a white-noise-like component to the power spectrum and can therefore bias the measurement of diffuse ICM distributions — are identified using \texttt{wavdetect}. For each detected source, we construct a circular exclusion region centered on the source position, with radius equal to the local $r_{90}$, i.e. the radius enclosing 90$\%$ of the encircled energy fraction. The PSF map is generated with \texttt{mkpsfmap} assuming ${\rm ECF}=0.9$, and therefore naturally accounts for the degradation of the Chandra PSF with off-axis angle. The resulting masks are applied prior to the Fourier analysis, ensuring that only the diffuse intracluster emission contributes to the measured power spectrum and coherence. Nevertheless, we verified that the application of the mask has a negligible impact on the \textsl{coherence length} measurement, as it primarily affects scales much smaller than the \textsl{coherence length}. We show the resulting point-source mask for MACSJ0416.1$-$2403 in Fig~\ref{MACSJ0416_mask}. This cluster was selected as an illustrative case because it has one of the largest fields of view in our sample, allowing the masking procedure to be clearly visualized across a wide range of off-axis angles.

The final products of this procedure are background-subtracted, exposure-corrected Chandra images, together with the corresponding masks, which form the basis for the subsequent power-spectrum and coherence analysis. As discussed in Section~\ref{results}, Figures~\ref{ABELL2261_coherence}, \ref{MACSJ0416.1-2403_coherence} (with the corresponding point-source mask shown in Fig.~\ref{MACSJ0416_mask}, for the largest map among the three cases), and \ref{CLJ0152.7-1357_coherence} illustrate representative examples drawn from the CLASH, HSTFF, and RELICS samples, respectively.

\section{Gas-Mass coherence} \label{coherence}

From the lensing mass maps and their associated error maps described in subsection~\ref{lens_data}, as well as from the exposure-corrected, background-subtracted, masked X-ray surface-brightness maps $\tfrac{1}{2}(A+B)$ and their corresponding noise maps $\tfrac{1}{2}(A-B)$ described in subsection~\ref{xray}, we compute the Fourier transforms using the discrete FFT:

\begin{equation}
\Delta_m (\mathbf{q})=\int F_m(x)\exp(-i\mathbf{x}\cdot\mathbf{q})d^2x,
\end{equation}
\begin{equation}
\Delta_X (\mathbf{q})=\int F_X(x)\exp(-i\mathbf{x}\cdot\mathbf{q})d^2x,
\end{equation}

where $F_m(x)$ and $F_X(x)$ denote, respectively, the lensing-derived mass (or mass-error) maps and the X-ray signal (or noise) maps, and $\mathbf{x}$ is the real-space coordinate and $\mathbf{q}=2\pi\mathbf{k}$ the wave vector, with $|\mathbf{k}|=1/\theta$ and $\theta$ the angular scale. In our previous work (~\citealt{https://Cerini22},~\citealt{Cerini2025}), we computed fluctuation fields by subtracting the spatial averages $\langle F_m \rangle$ and $\langle F_X \rangle$ from the corresponding maps, following the standard fluctuation-analysis approach. However, in the present context — where we analyze clusters spanning a wide range of dynamical states — modeling and subtracting a global smooth component (e.g., a $\beta$-model profile; \citealt{Churazov_2012},~\citealt{zhuravleva17}) is not appropriate, as such profiles do not provide an adequate description for systems out of equilibrium. In addition, subtracting a constant offset does not affect the power spectrum or the coherence, as these statistics are insensitive to the zero-frequency (mean) component. For this reason, we retain the full fields without mean subtraction, also to fully characterize the multi-scale structure of the underlying mass and X-ray distributions.

The 1D auto-power spectra are

\begin{equation}
P_m(q)=\langle |\Delta_m (\mathbf{q})|^2 \rangle,
\end{equation}
\begin{equation}
P_X(q)=\langle |\Delta_X (\mathbf{q})|^2 \rangle,
\end{equation}

where the average is taken over all Fourier modes in the annulus $[q,q+dq]$.

From these quantities, we compute the auto-power spectra associated with the signal and noise components of each dataset. For the lensing maps, we obtain the mass-data power spectrum $P^{data}_m(q)$ and the mass-error power spectrum $P^{err}_m(q)$. Similarly, for the X-ray maps we compute the power spectrum of the signal map, $P^{data}_X(q)$, derived from $\tfrac{1}{2}(A+B)$, and the noise power spectrum $P^{err}_X(q)$, derived from $\tfrac{1}{2}(A-B)$.
The “clean’’ auto-power spectra for the mass and X-ray fields are then obtained by subtracting the corresponding noise contributions:

\begin{equation}
P_m^{\mathrm{clean}}(q) = P_m^{\mathrm{data}}(q) - P_m^{\mathrm{err}}(q),
\end{equation}

\begin{equation}
P_X^{\mathrm{clean}}(q) = P_X^{\mathrm{data}}(q) - P_X^{\mathrm{err}}(q).
\end{equation}

Uncertainties on the clean power spectra are propagated from the individual terms using standard error-propagation rules.

The cross-power spectrum is

\begin{IEEEeqnarray}{rCl}
P^{clean}_{mX}(q) &=& 
\langle \Delta^{\mathrm{data}}_{m}(q)\,
         \Delta^{\mathrm{data}}_{X}{}^{*}(q) \rangle \\
&=& \Re^{\mathrm{data}}_{m}(q)\,
    \Re^{\mathrm{data}}_{X}(q) \nonumber\\
&& {}+\, \Im^{\mathrm{data}}_{m}(q)\,
        \Im^{\mathrm{data}}_{X}(q). \nonumber
\end{IEEEeqnarray}

Because the noise in both the mass-error maps and the X-ray difference maps is uncorrelated with the astrophysical signal and with each other, their expectation value in the cross-power is zero. Thus, only the real and imaginary parts of the data maps enter the estimator for $P_{mX}(q)$.

The coherence is defined as

\begin{equation}
C(q)=(P^{clean}_{mX})^2(q)/[P^{clean}_m(q)P^{clean}_X(q)],
\label{c_formula}
\end{equation}

and quantifies, in Fourier space, how well the gas traces the gravitational potential at each scale. Values of $C\simeq1$ indicate nearly perfect correlation, while $C\simeq0$ corresponds to uncorrelated structure. As shown in \citealt{https://Cerini22}, this quantity is highly sensitive to departures from equilibrium and reveals spatial structures not easily identifiable in real space.

Our dynamical state indicator is the \textsl{coherence length} $\ell_{\rm CR}$, defined as the smallest scale at which $C > 0.9$. Relaxed clusters exhibit high coherence at all scales and therefore small $\ell_{\rm CR}$, whereas dynamically disturbed systems show suppressed coherence at small and intermediate scales and correspondingly larger $\ell_{\rm CR}$.

The statistical uncertainties on the auto- and cross-power spectra are given by

\begin{equation}
\sigma_{P^{clean}_m}=\frac{P^{clean}_m(q)}{\sqrt{0.5N_q}},
\end{equation}
\begin{equation}
\sigma_{P^{clean}_X}=\frac{P^{clean}_X(q)}{\sqrt{0.5N_q}},
\end{equation}
\begin{equation}
\sigma_{P^{clean}_{mX}}(q) =
\sqrt{\frac{P^{clean}_{mX}(q)^2 + \tfrac{1}{2} P^{clean}_m(q)\,P^{clean}_X(q)}{N_q}},
\end{equation}

with $N_q/2$ independent Fourier modes per annulus. 
The full expression for the cross-power uncertainty is required because the commonly used approximation $\sigma_{P_{mX}}=\sqrt{P_m(q)P_X(q)/N_q}$ assumes that the Fourier modes of the two fields are effectively independent. This assumption does not hold for DM and ICM distributions in galaxy clusters. Indeed, at large and intermediate spatial scales, the DM and ICM distributions are substantially correlated, and in relaxed systems this correlation can remain high across essentially all scales. As a consequence, the variance of the cross term includes an additional contribution from the intrinsic scatter of the correlated signal, proportional to $P_{mX}^2$, such that $\mathrm{Var}[\hat P_{mX}] \propto P_m P_X + P_{mX}^2$ (e.g. \citealt{Cabré2007}, \citealt{Tristram2005}). Neglecting this term leads to a systematic underestimation of the uncertainty, particularly in the high-coherence regime where the two fields closely trace each other.
This choice differs from previous works on cosmic background fluctuations (e.g. \citealt{Kashlinsky_2018}), where the approximation based solely on the product of the auto-power spectra is adopted. In those cases, the signals being cross-correlated (e.g. CIB and CXB) exhibit a weaker correlation at all scales. In contrast, for galaxy clusters the strong physical coupling between DM and ICM is non-negligible and adding the term $P_{mX}^2$ is essential for a reliable uncertainty estimate. For these reasons, we adopt the complete expression, which provides an unbiased and numerically stable estimate of the cross-power uncertainty for both relaxed and dynamically disturbed systems.

Because the coherence is highly nonlinear and bounded ($0\le C\le1$), errors are estimated via the Fisher transformation (\citealt{Fisher1915}; \citealt{Kashlinsky_2018}). Writing $R=\sqrt{C}$, we define

\begin{equation}
Z=\frac{1}{2}\ln{\left(\frac{1+R}{1-R}\right)},
\end{equation}

which is approximately normally distributed. Confidence intervals in $C$ are recovered through the inverse relation $C=(\tanh Z)^2$. The corresponding variance is

\begin{IEEEeqnarray}{rCl}
\sigma_Z^2 &=& \frac{C_0}{(1-C_0)^2}
\Bigg(
\frac{\sigma_{P^{\mathrm{clean}}_{mX}}^2}{\big(P^{\mathrm{clean}}_{mX}\big)^2}
+ \frac{1}{4}\frac{\sigma_{P^{\mathrm{clean}}_{m}}^2}{\big(P^{\mathrm{clean}}_{m}\big)^2} \nonumber\\
&& \qquad
+ \frac{1}{4}\frac{\sigma_{P^{\mathrm{clean}}_{X}}^2}{\big(P^{\mathrm{clean}}_{X}\big)^2}
\Bigg).
\end{IEEEeqnarray}

In this work we compute the error on the \textsl{coherence length} as

\begin{equation}
\sigma_{\ell_{\rm CR}} = \left(\frac{\sigma_C(\ell_{\rm CR})}{C(\ell_{\rm CR})}\right)\ell_{\rm CR}.
\end{equation}

In our previous work (\citealt{Cerini2025}), we adopted the standard error propagation $\sigma_{\ell_{\rm CR}}^2=\sigma_C^2 \big/ \left(\left.\frac{dC}{dl}\right|_{l=\ell_{\rm CR}}\right)^2$. In the present analysis, we instead adopt the proportional form given above in order to avoid overestimating the uncertainty on the coherence length. This choice is motivated by the fact that the uncertainty on the coherence is derived from the Fisher transformation, as described above, and therefore does not strictly satisfy the assumptions underlying standard error propagation based on local linearization. In particular, near the threshold value $C\simeq0.9$, the coherence function becomes progressively flatter, and its inverse correspondingly steeper. As a result, the derivative term $\left(dC/dl\right)^{-1}$ can become artificially large, leading to inflated and unstable error estimates on $\ell_{\rm CR}$. For this reason, we directly propagate the fractional uncertainty on the coherence evaluated at $\ell_{\rm CR}$, which provides a more robust and stable estimate of the uncertainty on the coherence length.

To mitigate edge-induced spectral leakage associated with the finite extent of the maps, we apodize the images with a Hann window prior to the Fourier transform, consistently with what we did in~\citealt{Cerini2025}. Similar choices have been explicitly adopted, for instance, in patch-based CMB power-spectrum analyses of finite fields, as well as in 21\,cm power-spectrum studies (e.g.~\citealt{Coble2003}, \citealt{Joachimi2011}, \citealt{Munshi2024}).

For a consistent Fourier analysis, it is essential that the maps being compared share the same field of view, pixel scale, and grid geometry. To ensure this, we reproject the $\kappa$ and X-ray maps onto a common reference grid, matching both the map extent and pixel sampling. The reprojection is performed using the \texttt{reproject} package in \texttt{astropy}, always matching to the coarser of the two datasets to avoid introducing artificial resolution. In addition, as discussed in Section~\ref{lensing_reconstruction}, we restrict the sample of lensing models to those with angular resolution comparable to on-axis Chandra observations, discarding models with significantly coarser resolution. This ensures that the \textsl{coherence length} can be robustly measured down to small spatial scales, particularly for the most relaxed clusters. As a result, all maps used in this analysis have a similar effective angular resolution of $\sim1''$.

\startlongtable
\begin{deluxetable*}{l c c c c l c c}
\tablewidth{0pt}
\tabletypesize{\scriptsize}
\tablecaption{List of the 49 galaxy clusters analyzed in this work. Cluster names follow the HST MAST archive convention. For each system, we report the parent survey, redshift, J2000 coordinates (from the Chandra archive), and the coherence length measured for each available lensing model, both in physical units and normalized by $r_{500}$. The values of $r_{500}$ are adopted from the reference works describing the HSTFF, CLASH, and RELICS cluster samples (e.g.~\citealt{Lotz_2017},~\citealt{Postman_2012} and~\citealt{Coe_2019}). \label{table:clusters_models}}
\tablehead{
\colhead{Cluster Name} &
\colhead{$z$} &
\colhead{Sample} &
\colhead{RA (J2000, deg)} &
\colhead{DEC (J2000, deg)} &
\colhead{Model} &
\colhead{$\ell_{\rm CR}$ (kpc)} &
\colhead{$\ell_{\rm CR}/r_{500}$}
}
\startdata
Abell 209 & 0.206 & CLASH & 23.018 & -13.564 & zitrin-ltm-gauss\_v2 & $684.27 \pm 135.97$ & $0.49 \pm 0.10$ \\
& & & & & zitrin-nfw\_v2 & $684.27 \pm 135.97$ & $0.49 \pm 0.10$ \\[2pt]
Abell 370 & 0.375 & HSTFF & 39.964 & -1.592 & cats\_v4 & $620.61 \pm 120.26$ & $0.46 \pm 0.09$ \\[2pt]
Abell 383 & 0.187 & CLASH & 42.011 & -3.473 & zitrin-ltm-gauss\_v2 & $102.69 \pm 19.98$ & $0.07 \pm 0.015$ \\
& & & & & zitrin-ltm\_v1\_z2.55 & $91.64 \pm 16.98$ & $0.07 \pm 0.01$ \\
& & & & & zitrin-nfw\_v2 & $198.19 \pm 38.57$ & $0.14 \pm 0.03$ \\[2pt]
Abell 697 & 0.282 & RELICS & 130.738 & 36.419 & glafic\_v2 & $340.22 \pm 64.22$ & $0.24 \pm 0.04$ \\
& & & & & zitrin-ltm-gauss\_v2 & \nodata & \nodata \\[2pt]
Abell 1758 & 0.280 & RELICS & 203.132 & 50.519 & glafic\_v2 & $1079.47 \pm 194.39$ & $0.83 \pm 0.15$ \\[2pt]
Abell 1763 & 0.228 & RELICS & 203.849 & 40.956 & glafic\_v2 & $554.09 \pm 75.92$ & $0.42 \pm 0.06$ \\[2pt]
Abell 2163 & 0.203 & RELICS & 243.890 & -6.190 & glafic\_v2 & \nodata & \nodata \\
& & & & & lenstool\_v1 & \nodata & \nodata \\[2pt]
Abell 2261 & 0.224 & CLASH & 260.666 & 32.164 & zitrin-ltm-gauss\_v2 & $164.07 \pm 31.93$ & $0.10 \pm 0.02$ \\
& & & & & zitrin-nfw\_v2 & $164.07 \pm 31.93$ & $0.10 \pm 0.02$ \\[2pt]
Abell 2537 & 0.297 & RELICS & 347.080 & -2.166 & glafic\_v2 & $185.81 \pm 35.07$ & $0.16 \pm 0.03$ \\
& & & & & lenstool\_v1 & $205.02 \pm 39.93$ & $0.18 \pm 0.03$ \\[2pt]
Abell 2744 & 0.308 & HSTFF & 3.598 & -30.398 & cats\_v4.1 & $2367.28 \pm 322.24$ & $1.42 \pm 0.19$ \\[2pt]
Abell 2813 & 0.292 & RELICS & 10.884 & -20.598 & lenstool\_v1 & $513.76 \pm 104.78$ & $0.40 \pm 0.08$ \\[2pt]
Abell 3192 & 0.425 & RELICS & 59.744 & -29.913 & glafic\_v2 & $842.27 \pm 115.41$ & $0.72 \pm 0.10$ \\
& & & & & lenstool\_v1 & $654.11 \pm 133.40$ & $0.56 \pm 0.11$ \\[2pt]
Abell S295 & 0.300 & RELICS & 41.352 & -53.035 & glafic\_v2 & $674.06 \pm 92.36$ & $0.56 \pm 0.08$ \\
& & & & & zitrin-ltm-gauss\_v2 & $532.47 \pm 108.95$ & $0.44 \pm 0.09$ \\[2pt]
Abell S1063 & 0.348 & HSTFF & 342.194 & -44.492 & cats\_v4.1 & $688.05 \pm 127.10$ & $0.46 \pm 0.08$ \\
& & & & & diego\_v4.1 & $809.58 \pm 112.76$ & $0.54 \pm 0.07$ \\
& & & & & keeton\_v4 & $744.40 \pm 140.50$ & $0.49 \pm 0.09$ \\
& & & & & sharon\_v4 & $578.11 \pm 117.90$ & $0.38 \pm 0.08$ \\
& & & & & williams\_v4.1 & $745.30 \pm 140.63$ & $0.49 \pm 0.09$ \\
& & & & & zitrin-ltm-gauss\_v1 & $711.56 \pm 96.63$ & $0.47 \pm 0.06$ \\
& & & & & zitrin-ltm\_v1 & $711.56 \pm 96.63$ & $0.47 \pm 0.06$ \\
& & & & & zitrin-nfw\_v1 & $711.56 \pm 96.63$ & $0.47 \pm 0.06$ \\[2pt]
ACT-CLJ0102-49151 & 0.870 & RELICS & 15.710 & -49.304 & glafic\_v3 & $930.51 \pm 180.61$ & $0.83 \pm 0.16$ \\[2pt]
CLJ0152.7-1357 & 0.833 & RELICS & 28.205 & -13.939 & glafic\_v2 & $1139.09 \pm 155.86$ & $1.26 \pm 0.17$ \\
& & & & & lenstool\_v1 & $891.17 \pm 181.75$ & $0.98 \pm 0.20$ \\[2pt]
CLJ1226+3332 & 0.890 & CLASH & 186.744 & 33.574 & zitrin-ltm-gauss\_v2 & $944.16 \pm 268.81$ & $0.75 \pm 0.21$ \\
& & & & & zitrin-nfw\_v2 & $1481.77 \pm 268.81$ & $1.18 \pm 0.21$ \\[2pt]
MACSJ0025.4-1222 & 0.586 & RELICS & 6.413 & -12.405 & glafic\_v2 & \nodata & \nodata \\
& & & & & zitrin-ltm-gauss\_v1 & $942.76 \pm 186.62$ & $0.84 \pm 0.17$ \\[2pt]
MACSJ0035.4-2015 & 0.352 & RELICS & 8.865 & -20.299 & glafic\_v2 & $208.31 \pm 39.32$ & $0.17 \pm 0.03$ \\
& & & & & lenstool\_v1 & $136.76 \pm 24.70$ & $0.11 \pm 0.02$ \\[2pt]
MACSJ0159.8-0849 & 0.405 & RELICS & 29.975 & -8.801 & glafic\_v2 & $227.36 \pm 42.91$ & $0.19 \pm 0.04$ \\[2pt]
MACSJ0329-02 & 0.450 & CLASH & 52.452 & -2.209 & zitrin-ltm-gauss\_v2 & $215.77 \pm 38.97$ & $0.19 \pm 0.03$ \\
& & & & & zitrin-nfw\_v2 & \nodata & \nodata \\[2pt]
MACSJ0416.1-2403 & 0.397 & HSTFF & 64.007 & -24.062 & caminha\_v4 & $746.21 \pm 200.60$ & $0.53 \pm 0.14$ \\
& & & & & cats\_v4.1 & $746.21 \pm 200.60$ & $0.53 \pm 0.14$ \\
& & & & & diego\_v4.1 & $734.92 \pm 98.77$ & $0.52 \pm 0.07$ \\
& & & & & keeton\_v4 & $587.70 \pm 110.93$ & $0.42 \pm 0.08$ \\
& & & & & sharon\_v4 & $628.71 \pm 128.22$ & $0.45 \pm 0.09$ \\[2pt]
MACSJ0417.5-1154 & 0.443 & RELICS & 64.393 & -11.947 & glafic\_v3 & $455.21 \pm 85.92$ & $0.33 \pm 0.06$ \\
& & & & & lenstool\_v2 & \nodata & \nodata \\[2pt]
MACSJ0429-02 & 0.399 & CLASH & 67.409 & -2.922 & zitrin-ltm-gauss\_v2 & $147.94 \pm 26.72$ & $0.12 \pm 0.02$ \\
& & & & & zitrin-nfw\_v2 & $274.13 \pm 49.52$ & $0.22 \pm 0.04$ \\[2pt]
MACSJ0553.4-3342 & 0.352 & RELICS & 88.387 & -33.720 & lenstool\_v1 & \nodata & \nodata \\[2pt]
MACSJ0717.5+3745 & 0.545 & HSTFF & 109.435 & 37.764 & cats\_v4.1 & $1970.20 \pm 404.53$ & $1.16 \pm 0.24$ \\[2pt]
MACSJ0744.9+3927 & 0.698 & RELICS & 116.224 & 39.495 & zitrin-ltm-gauss\_v2 & $490.89 \pm 88.67$ & $0.47 \pm 0.08$ \\
& & & & & zitrin-nfw\_v2 & $1128.68 \pm 230.23$ & $1.08 \pm 0.22$ \\[2pt]
MACSJ1115.9+0129 & 0.355 & CLASH & 168.984 & 1.531 & zitrin-ltm-gauss\_v2 & $435.64 \pm 84.77$ & $0.35 \pm 0.07$ \\
& & & & & zitrin-nfw\_v2 & $313.59 \pm 61.02$ & $0.25 \pm 0.05$ \\[2pt]
MACSJ1149.5+2223 & 0.544 & HSTFF & 177.404 & 22.459 & cats\_v4.1 & $886.30 \pm 275.59$ & $0.52 \pm 0.16$ \\
& & & & & diego\_v4.1 & $876.46 \pm 117.89$ & $0.52 \pm 0.07$ \\
& & & & & keeton\_v4 & $961.54 \pm 131.75$ & $0.57 \pm 0.08$ \\[2pt]
MACSJ1206.2-0847 & 0.439 & CLASH & 181.567 & -8.767 & zitrin-ltm-gauss\_v2 & $290.09 \pm 52.40$ & $0.23 \pm 0.04$ \\
& & & & & zitrin-nfw\_v2 & $290.09 \pm 52.40$ & $0.23 \pm 0.04$ \\[2pt]
MACSJ1311.0-0310 & 0.494 & RELICS & 197.770 & -3.153 & zitrin-ltm-gauss\_v2 & $420.92 \pm 76.03$ & $0.43 \pm 0.08$ \\
& & & & & zitrin-nfw\_v2 & $572.97 \pm 197.41$ & $0.58 \pm 0.20$ \\[2pt]
MACSJ1423.8+2404 & 0.545 & CLASH & 215.902 & 24.047 & zitrin-ltm-gauss\_v2 & $289.99 \pm 56.43$ & $0.28 \pm 0.05$ \\
& & & & & zitrin-nfw\_v2 & $777.46 \pm 221.35$ & $0.74 \pm 0.21$ \\[2pt]
MACSJ1532.9+3021 & 0.363 & CLASH & 233.244 & 30.350 & zitrin-ltm-gauss\_v2 & $132.51 \pm 24.63$ & $0.10 \pm 0.02$ \\
& & & & & zitrin-nfw\_v2 & $198.77 \pm 36.94$ & $0.15 \pm 0.03$ \\[2pt]
MACSJ1621.3+3810 & 0.463 & RELICS & 245.383 & 38.139 & glafic\_v3 & $603.11 \pm 116.94$ & $0.48 \pm 0.09$ \\[2pt]
MACSJ1720.3+3536 & 0.391 & RELICS & 260.101 & 35.579 & zitrin-ltm-gauss\_v2 & $646.34 \pm 125.78$ & $0.53 \pm 0.10$ \\
& & & & & zitrin-nfw\_v2 & $465.26 \pm 90.54$ & $0.38 \pm 0.07$ \\[2pt]
MACSJ1931.8-2635 & 0.352 & CLASH & 292.963 & -26.603 & zitrin-ltm-gauss\_v2 & $225.63 \pm 43.91$ & $0.19 \pm 0.04$ \\
& & & & & zitrin-nfw\_v2 & $604.89 \pm 172.22$ & $0.52 \pm 0.15$ \\[2pt]
MACSJ2129.4-0741 & 0.589 & CLASH & 322.330 & -7.659 & zitrin-ltm-gauss\_v2 & $297.62 \pm 55.08$ & $0.24 \pm 0.04$ \\
& & & & & zitrin-nfw\_v2 & $446.43 \pm 82.61$ & $0.36 \pm 0.07$ \\[2pt]
MACSJ2214.9-1359 & 0.502 & RELICS & 333.750 & -14.040 & glafic\_v3 & $506.33 \pm 97.77$ & $0.40 \pm 0.08$ \\[2pt]
MS1008.1-1224 & 0.306 & RELICS & 152.673 & -12.655 & lenstool\_v1 & \nodata & \nodata \\[2pt]
PLCK-G287.0+32.9 & 0.540 & RELICS & 177.702 & -28.101 & glafic\_v2 & $787.53 \pm 107.91$ & $0.52 \pm 0.07$ \\
& & & & & zitrin-ltm-gauss\_v1 & $1025.40 \pm 202.80$ & $0.68 \pm 0.13$ \\[2pt]
PLCK-G171.9-40.7 & 0.270 & RELICS & 48.239 & 8.409 & glafic\_v2 & $329.87 \pm 62.26$ & $0.23 \pm 0.04$ \\
& & & & & zitrin-ltm-gauss\_v1 & $292.71 \pm 77.30$ & $0.21 \pm 0.05$ \\[2pt]
PLCK-G004.5-19.5 & 0.540 & RELICS & 289.297 & -33.554 & lenstool\_v1 & $1360.79 \pm 237.90$ & $1.07 \pm 0.19$ \\[2pt]
RXJ2129.7+0005 & 0.234 & CLASH & 322.426 & 0.036 & zitrin-ltm-gauss\_v2 & $169.51 \pm 32.99$ & $0.14 \pm 0.03$ \\
& & & & & zitrin-nfw\_v2 & $454.44 \pm 149.86$ & $0.38 \pm 0.12$ \\[2pt]
RXCJ0142.9+4438 & 0.341 & RELICS & 25.662 & 44.610 & glafic\_v2 & $281.08 \pm 53.05$ & $0.21 \pm 0.04$ \\
& & & & & lenstool\_v1 & $225.02 \pm 43.82$ & $0.17 \pm 0.03$ \\[2pt]
RXCJ0232.2-4420 & 0.284 & RELICS & 38.057 & -44.297 & lenstool\_v1 & $503.98 \pm 102.78$ & $0.40 \pm 0.08$ \\[2pt]
RXJ1347-1145 & 0.451 & CLASH & 206.863 & -11.788 & zitrin-ltm-gauss\_v2 & $398.59 \pm 72.18$ & $0.29 \pm 0.05$ \\
& & & & & zitrin-nfw\_v2 & $739.58 \pm 133.92$ & $0.54 \pm 0.10$ \\[2pt]
RXCJ0600.1-2007 & 0.460 & RELICS & 89.971 & -20.118 & glafic\_v3 & $464.49 \pm 87.67$ & $0.35 \pm 0.07$ \\
& & & & & lenstool\_v1 & \nodata & \nodata \\[2pt]
SMACSJ0723.3-7327 & 0.390 & RELICS & 110.802 & -73.509 & glafic\_v2 & $580.77 \pm 109.62$ & $0.46 \pm 0.09$ \\
& & & & & lenstool\_v1 & $621.30 \pm 126.71$ & $0.50 \pm 0.10$ \\
& & & & & zitrin-ltm-gauss\_v2 & $164.50 \pm 31.31$ & $0.13 \pm 0.02$ \\[2pt]
SPT-CLJ0615-5746 & 0.972 & RELICS & 94.032 & -57.742 & glafic\_v2 & \nodata & \nodata \\
& & & & & lenstool\_v1 & $551.22 \pm 189.86$ & $0.59 \pm 0.20$ \\[2pt]
\enddata
\end{deluxetable*}

\begin{figure*}
\centering
\includegraphics[width=1.00\textwidth]{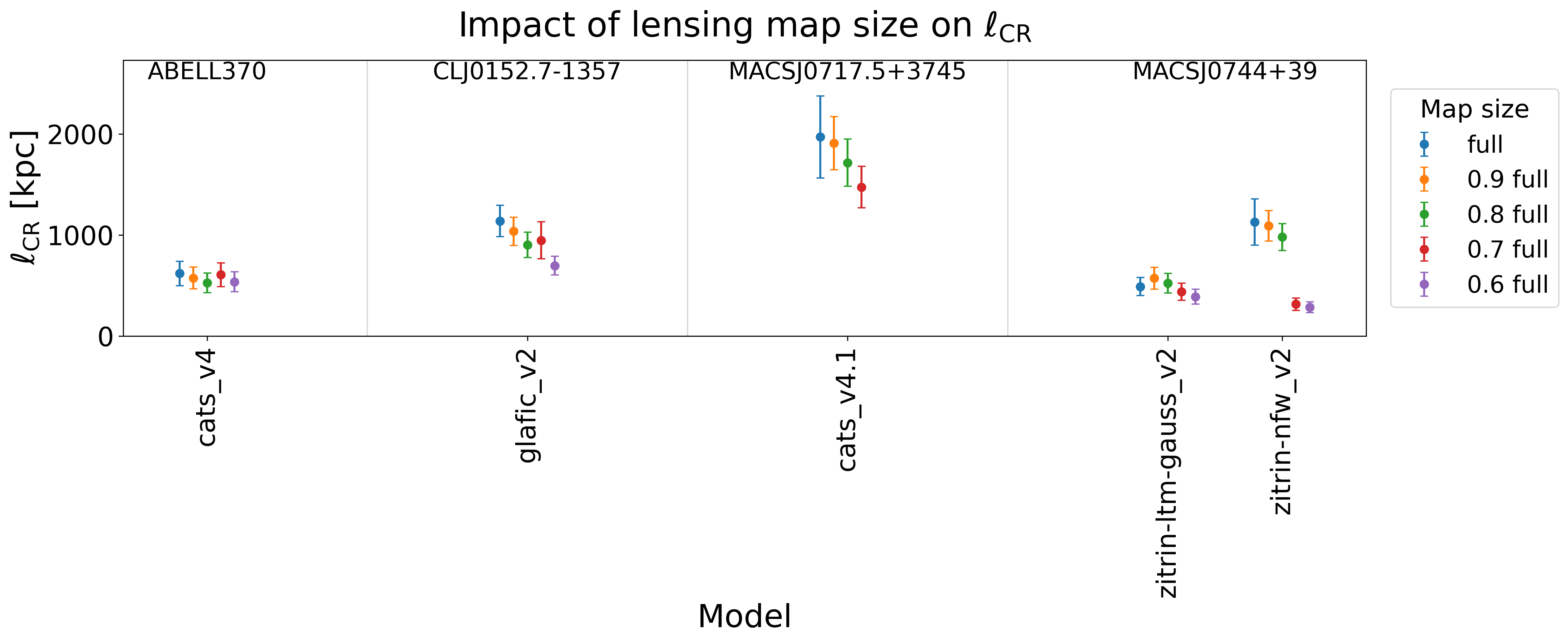}
\caption{\textsl{Coherence length}, $\ell_{\rm CR}$, measured from progressively smaller centered cutouts of the original convergence maps for four clusters selected to have the largest available lensing map extents. Different colors correspond to different map sizes, normalized to the full map extent. The figure illustrates the impact of map size on the inferred \textsl{coherence length.}}
\label{size_effect}
\end{figure*}

\section{Results} \label{results}
\subsection{Coherence Behavior Across Different Dynamical States}\label{examples_c}

We first examine the behavior of the DM-gas coherence using three representative clusters that span the range of dynamical states probed in this work. Figures 3, 4, and 6 show examples from the CLASH, HSTFF, and RELICS samples, respectively. These systems illustrate the expected progression from high coherence on small scales in dynamically regular clusters to a loss of coherence on increasingly larger scales in disturbed and merging systems.

For the relaxed CLASH cluster ABELL 2261, the coherence curves obtained from the available lensing mass reconstructions are nearly identical over the full range of scales. We measure a tightly constrained coherence length of $\ell_{\rm CR}=164\pm32$ kpc, corresponding to $\ell_{\rm CR}/r_{500}=0.105\pm0.020$. This small value indicates that the lensing-derived mass distribution and the X-ray-emitting ICM remain highly coherent down to small physical scales, as expected for a dynamically regular and relaxed system.

MACSJ0416.1$-$2403 provides an intermediate case. This HSTFF cluster is known to be dynamically complex, and our analysis yields substantially larger coherence lengths, spanning $\ell_{\rm CR}\simeq 590$--$750$ kpc, or $\ell_{\rm CR}/r_{500}\simeq0.42$--$0.53$, depending on the adopted lensing model. The broader transition in the coherence signal reflects the presence of scale-dependent DM-gas misalignment and complexity parlayed by the presence of spatial substructures. The scatter between lensing reconstructions also indicates that, for morphologically complex clusters, the inferred coherence length is sensitive to specific details of the mass model.

The most disturbed example considered here is CLJ0152.7$-$1357. For this high-redshift merging RELICS cluster, we measure large coherence lengths, $\ell_{\rm CR}\simeq 891$--$1139$ kpc, corresponding to $\ell_{\rm CR}/r_{500}\simeq0.98$--$1.26$. The coherence therefore drops below the adopted threshold only on large physical scales, consistent with a system in which the ICM and the collisionless mass distribution are substantially displaced by merger activity. This interpretation agrees with previous X-ray studies that identify CLJ0152.7$-$1357 as a massive, multi-component merging system.

Together, these examples demonstrate that $\ell_{\rm CR}$ provides a physically interpretable, scale-resolved diagnostic of cluster dynamical state. Small coherence lengths identify systems in which the gas and mass distributions remain aligned on small scales, while large coherence lengths mark clusters in which merger-driven disturbances, substructure, or asymmetries drive significant DM-ICM decorrelation. They also show that the lensing reconstruction itself can introduce measurable scatter, motivating the sample-wide analysis results outlined below.

\subsection{Distribution and robustness of the coherence length}

We next quantify the distribution of coherence lengths across the full sample of 49 clusters. For each cluster, we compute the mean normalized coherence length, $\langle \ell_{\rm CR}/r_{500}\rangle$, across all available lensing models, together with the model-to-model scatter, $\sigma_{\rm models}$. We also compute the relative scatter, $\sigma_{\rm models}/\langle \ell_{\rm CR}/r_{500}\rangle$, as a dimensionless measure of the robustness of the inferred coherence length. These quantities are summarized in Figure 7.

Adopting the threshold introduced in \citep{Cerini2025}, in which the most relaxed systems are identified by $\ell_{\rm CR}/r_{500}<0.2$, we find that only 9 out of 49 clusters satisfy $\langle \ell_{\rm CR}/r_{500}\rangle<0.2$. This corresponds to approximately 16\% of the sample. If the selection is broadened to include clusters with $\ell_{\rm CR}/r_{500}<0.4$, the relaxed or moderately regular fraction increases to 20 out of 49 systems, or approximately 41\%. The exact threshold is necessarily application-dependent: a stricter cut is better suited to scaling-relation work and precision cosmology, while a more permissive threshold may be appropriate for broader morphological classification and cluster assembly.

The sample exhibits substantial diversity in the robustness of the inferred coherence length. Only 18 of the 49 clusters show small model-to-model scatter, with $\sigma_{\rm models}/\langle \ell_{\rm CR}\rangle<0.2$, indicating consistent coherence measurements across lensing reconstructions. A further 9 clusters show significant scatter, with $0.2<\sigma_{\rm models}/\langle \ell_{\rm CR}\rangle<0.6$. In 16 clusters, a finite coherence length is obtained for only one lensing model, while the other reconstructions yield undefined values because the coherence does not reach the threshold $C=0.9$. Overall, 25 of the 49 systems, or approximately 52\% of the sample, show non-negligible model dependence.

This model dependence is an important empirical result. It indicates that the DM-gas coherence is sensitive not only to the physical dynamical state of the cluster, but also to the assumptions entering the lensing mass reconstruction, particularly when strong-lensing constraints dominate the mass model. Figure 8 illustrates this effect for SPT-CLJ0615$-$5746, where different convergence maps yield measurably different coherence behavior. This motivates future applications to homogeneous, model-independent weak-lensing reconstructions.

\subsection{Map-size effects and robust subsample selection}

The available HST lensing reconstructions are heterogeneous not only in methodology but also in spatial coverage. Although we restrict the analysis for each cluster to models with comparable map sizes, the physical extent of the convergence maps varies substantially across the full sample. Since Fourier-space coherence measurements are explicitly scale-dependent, this non-uniformity affects the range of physical scales accessible to the analysis and can influence the inferred coherence length.

A limited field of view can bias $\ell_{\rm CR}$ by excluding large-scale substructures and merger features. This effect is expected to be most important in disturbed systems, where the relevant DM-gas decorrelation often occurs at large radii. In such cases, truncating the map can artificially reduce the coherence length, making a cluster appear more relaxed than it would in a wider-field reconstruction. Thus, map size is a systematic limitation of the present heterogeneous sample.

To quantify this effect, we perform a controlled test using clusters with the largest available maps, selected to satisfy $L^{\kappa}{\rm map}/(2r{500})>1$. For each system, we recompute the coherence and coherence length after progressively reducing the map size using centered cutouts of the original convergence map. The results, shown in Figure 10, indicate that the coherence signal and inferred $\ell_{\rm CR}$ remain largely stable when the map is reduced to approximately 80\% of its full extent. Below this scale, deviations appear in some systems, most notably for the \texttt{zitrin-nfw v2} reconstruction of MACSJ0744+39.

We therefore adopt $L^{\kappa}{\rm map}/(2r{500})>0.8$ as a practical criterion for defining a robust subsample. Eleven clusters satisfy this requirement, and their coherence measurements are shown in Figure 11. This cut does not remove all sources of systematic uncertainty, but it identifies the systems for which the inferred coherence length is least likely to be dominated by field-of-view limitations.

\noindent\mbox{}\par
\subsection{Comparison with previous dynamical-state classifications}

We now compare the coherence-based results with previous dynamical-state classifications from the literature. This comparison is necessarily qualitative because existing classifications are based on heterogeneous diagnostics, including X-ray morphology, cool-core status, centroid shifts, optical structure, and radio evidence for mergers. A full quantitative cross-calibration of these indicators will be presented in future work.

The HSTFF clusters provide a useful consistency check because they are well known to be dynamically disturbed, strongly lensing systems. Our measurements reflect this expectation: all HSTFF clusters exhibit intermediate-to-large coherence lengths. The most extreme cases are MACS J0717+3745 and Abell 2744, two of the most structurally complex clusters in the HSTFF sample. We measure $\ell_{\rm CR}=1970\pm405$ kpc, or $1.16\pm0.24,r_{500}$, for MACS J0717+3745, and $\ell_{\rm CR}=2367\pm322$ kpc, or $1.42\pm0.19,r_{500}$, for Abell 2744. The remaining HSTFF systems occupy an intermediate range: MACS J1149.5+2223 spans $\ell_{\rm CR}\simeq876$--$962$ kpc, MACS J0416.1$-$2403 spans $\ell_{\rm CR}\simeq588$--$746$ kpc, Abell S1063 spans $\ell_{\rm CR}\simeq578$--$810$ kpc, and Abell 370 has $\ell_{\rm CR}=621\pm120$ kpc. These values are consistent with the known disturbed character of the HSTFF sample.

The RELICS sample spans a broader range of dynamical states. This diversity is reflected in the wider distribution of measured coherence lengths. Three RELICS clusters namely, Abell 2537, MACSJ0035.4$-$2015, and MACSJ0159.8$-$0849, have $\ell_{\rm CR}/r_{500}<0.2$ for all available lensing models. These systems are therefore identified by our method as dynamically regular, in agreement with previous studies that classify them as regular and/or cool-core clusters. For most other RELICS clusters, the coherence-based interpretation is broadly consistent with existing X-ray, optical, and radio classifications.

The main RELICS discrepancies occur for Abell 1758, MACSJ0744.9+3927, MACSJ1311.0$-$0310, MACSJ1621.3+3810, and RXCJ0232.2$-$4420. These five systems correspond to approximately 17\% of the RELICS subsample, have been classified in some previous work as cool-core clusters, yet in our analysis they show relatively large coherence lengths, with $\ell_{\rm CR}/r_{500}>0.4$. This suggests that DM--gas coherence may capture aspects of dynamical disturbance that are not fully encoded in cool-core versus non-cool-core classifications.

The tension is stronger for the CLASH subsample. CLASH clusters were selected to have comparatively regular X-ray morphologies and are predominantly cool-core systems. Nevertheless, we identify seven out of fourteen CLASH clusters with $\ell_{\rm CR}/r_{500}>0.4$, namely CLJ1226+3332, \allowbreak
MACSJ1115.9+0129, \allowbreak
MACSJ1206.2$-$0847, \allowbreak
MACSJ1423.8+2404, \allowbreak
MACSJ1931.8$-$2635, \allowbreak
RXJ2129.7+0005, and RXJ1347$-$1145.
These systems are classified as X-ray regular and/or cool-core in previous studies, but our coherence analysis indicates significant DM-ICM decorrelation. Abell 209 and MACSJ2129.4$-$0741 also have $\ell_{\rm CR}/r_{500}>0.4$, although earlier work reports higher ellipticities, larger centroid shifts, or non-cool-core classifications for these systems, making the discrepancy less severe.

Across the full sample, we find 12 discrepancies relative to previous classifications: 7 in CLASH and 5 in RELICS, corresponding to approximately 24\% of the 49 clusters analyzed here. In nearly all such cases, the coherence method classifies systems as less relaxed than standard X-ray morphological or cool-core diagnostics would suggest. This is qualitatively consistent with our previous TNG300 analysis, in which the coherence-based threshold identified a very large fraction of clusters as dynamically unrelaxed. The implication is that Fourier-space DM-gas coherence is a stringent dynamical-state diagnostic: it is sensitive to multi-scale misalignments between the collisionless mass component and the collisional ICM, rather than only to the central regularity of the X-ray emission.

Taken together, these results show that coherence analysis provides a complementary and physically motivated measure of the dynamical state of clusters. Its principal strength is that it directly combines the two dominant cluster components: the projected mass distribution, traced by gravitational lensing and dominated by dark matter, and the hot baryonic gas, traced by X-ray emission. The present application to HST and Chandra data demonstrates the feasibility of the method when applied to observed clusters, while also identifying the key systematics:(i) lensing-model dependence and (ii) non-uniform map size; issues that must be addressed with homogeneous weak-lensing datasets.

\section{Discussion and Conclusions} \label{conclusions}
In this work, we present the first application of the coherence analysis between gravitational-lensing reconstructed DM maps and X-ray observations of the ICM to a representative sample of observed galaxy clusters drawn from the HSTFF, CLASH, and RELICS programs. This study constitutes a fundamental step in the development of the method, following its initial introduction and validation on two observed clusters in~\citealt{https://Cerini22}, and its subsequent extension to a large statistical sample of simulated systems in the TNG300 suite in~\citealt{Cerini2025}.

The core of this framework is the measurement of the cross-correlation between the DM and gas distributions through the coherence, quantified via the cross- and auto-power spectra of the two fields. The main quantity used to characterize the dynamical state of the cluster is the \textsl{coherence length}, $\ell_{\rm CR}$, which traces the scale above which the two components remain highly correlated ($C \geq 0.9$). This provides a scale-resolved measure of the degree of alignment between the dominant gravitational component and the collisional baryonic gas.

The high angular resolution of both the lensing mass reconstructions from HST and the X-ray observations from Chandra enables us to probe the DM--gas correlation down to kpc scales across the cluster sample. In summary, the main results of this analysis can be outlined as follows:

\begin{itemize}
    \item The coherence-based method provides physically meaningful results, with lower \textsl{coherence lengths} generally associated with clusters exhibiting more regular morphologies, and larger values found in dynamically disturbed systems.

    \item We identify 9 clusters out of 49 (corresponding to $\sim16\%$ of the sample) with $<\ell_{\rm CR}>/r_{500} < 0.2$, which we interpret as the most relaxed systems. This threshold follows the definition adopted in~\citealt{Cerini2025}, where it was set by minimizing the scatter in the scaling relations. Extending the selection to $\ell_{\rm CR}/r_{500} < 0.4$, the number increases to 20 clusters ($\sim41\%$). We note that this threshold is to some extent arbitrary and depends on the specific goals of the dynamical state classification; in particular, a stricter threshold may be more appropriate for applications to scaling relations and precision cosmology.

    \item Across the full sample, we find a total of 12 discrepancies with respect to previous classifications (7 in CLASH and 5 in RELICS), corresponding to $\sim24\%$ of the clusters. In these cases, systems classified as X-ray regular and/or CC exhibit relatively large \textsl{coherence lengths} in our analysis. This suggests that the method has a tendency to highlight a larger population of dynamically unrelaxed systems compared to standard classifications.

    \item Despite the intrinsic robustness of the method, the lensing reconstruction introduces non-negligible scatter in the inferred \textsl{coherence length}. In particular, we find that in approximately $\sim52\%$ of the clusters the variation across models is significant, reflecting the sensitivity of the coherence signal to the details of the mass reconstruction, especially when strong-lensing information is involved.

    \item The cluster sample is heterogeneous also in terms of map size and spatial coverage. Our tests, based on comparing the coherence of the same clusters using different map sizes, show that reducing the map extent tends to decrease the inferred \textsl{coherence length}, thereby making clusters appear more relaxed. This effect is expected, as truncating the maps—particularly in dynamically disturbed systems—removes large-scale structures located at larger radii, while the central regions are typically more regular. Therefore, the inhomogeneity in map size across the sample is unlikely to be the primary driver of the discrepancies observed between our classification and more traditional ones, as it would bias the results in the opposite direction. Instead, in light of the results from the TNG300 simulations, these differences appear to be more closely related to the intrinsic sensitivity of the method to multi-scale misalignments and structural differences between the DM and gas distributions, as discussed above.

    \item A key strength of this method is that it provides a comprehensive multi-scale characterization of the dynamical state by directly combining the two main mass components, namely the DM distribution that dominates convergence maps, and the gas distribution traced by X-ray emission.
\end{itemize}

Building on this fundamental step, we are currently preparing the next stage of this analysis, aimed at applying the method to newly available SuperBIT data. In particular, following the recent public release of weak-lensing shape catalogs in \cite{saha2026}, we can construct convergence maps based exclusively on weak lensing. While these maps are intrinsically noisier, they are not based on any modeling assumptions, providing a more direct reconstruction of the projected mass distribution. This will constitute the first application of the coherence analysis to a homogeneous set of clusters with convergence maps derived from a single, consistent methodology. In addition, the larger field of view of the SuperBIT observations (of order $\sim0.1\,\mathrm{deg}^2$) will enable a more uniform map extent across the sample, mitigating the map-size systematics identified in the present work. This step is also essential for enabling a wider application of the method to large-area weak-lensing surveys such as Euclid and Roman, where homogeneous, model-independent mass reconstructions will be available for thousands of clusters.
This development will also be fundamental for extending the Fourier-based analysis beyond a pure dynamical-state investigation. Many previous Fourier-based studies --- from cosmic background analyses to investigations of ICM fluctuations --- have shown that complementary information can be extracted in Fourier space, which may be less apparent in real-space analyses. In this context, we are currently extending the framework to include additional tracers, like the galaxy distribution, with the goal of jointly characterizing the multi-component structure of galaxy clusters in a consistent and unified way. This will enable us to explore a broader range of physical processes, such as the role of feedback at different merger stages, while also opening new avenues for dark matter studies.

\begin{figure*}
\centering

\subfigure[]{\includegraphics[width=0.28\textwidth]{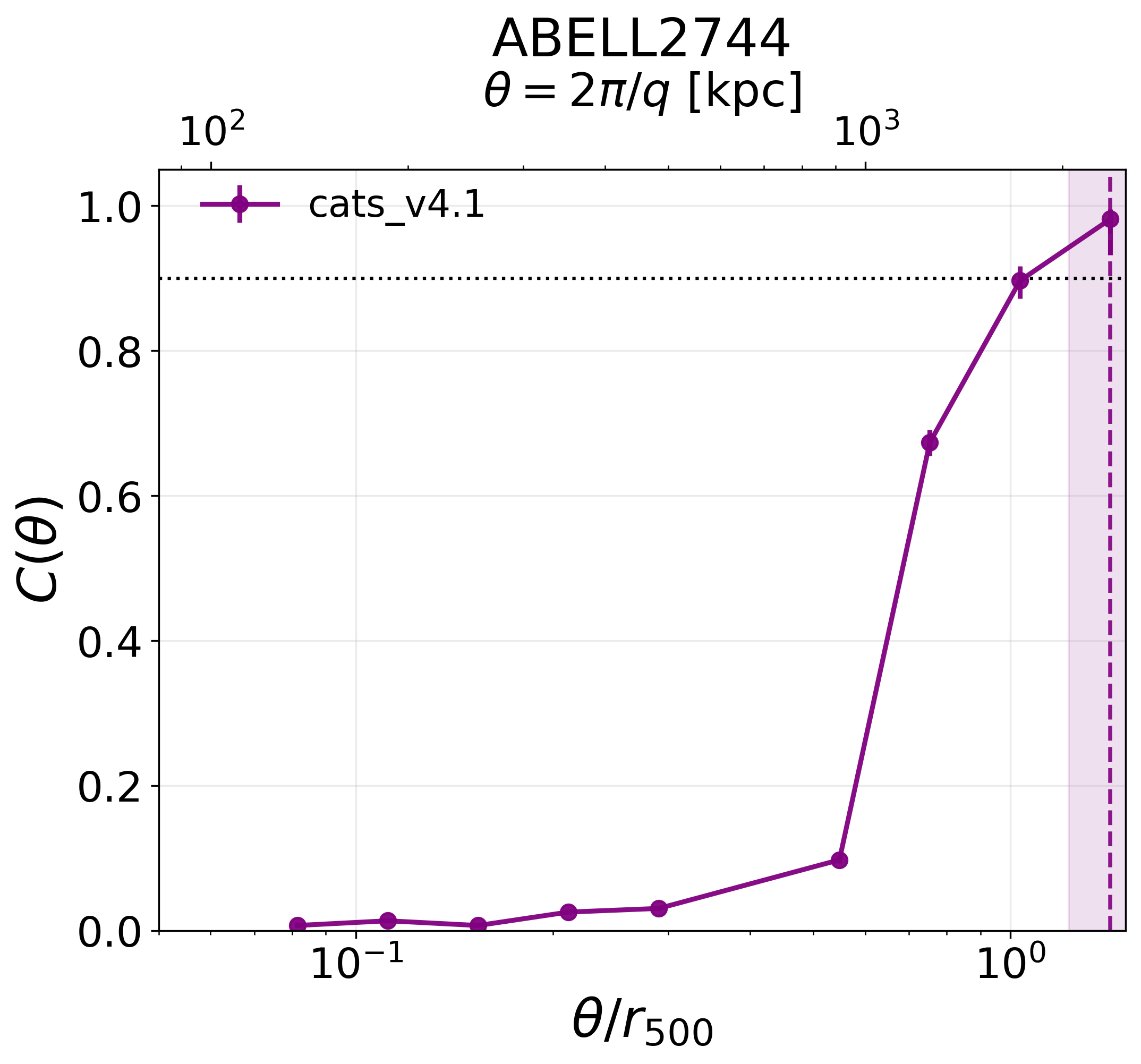}}
\subfigure[]{\includegraphics[width=0.28\textwidth]{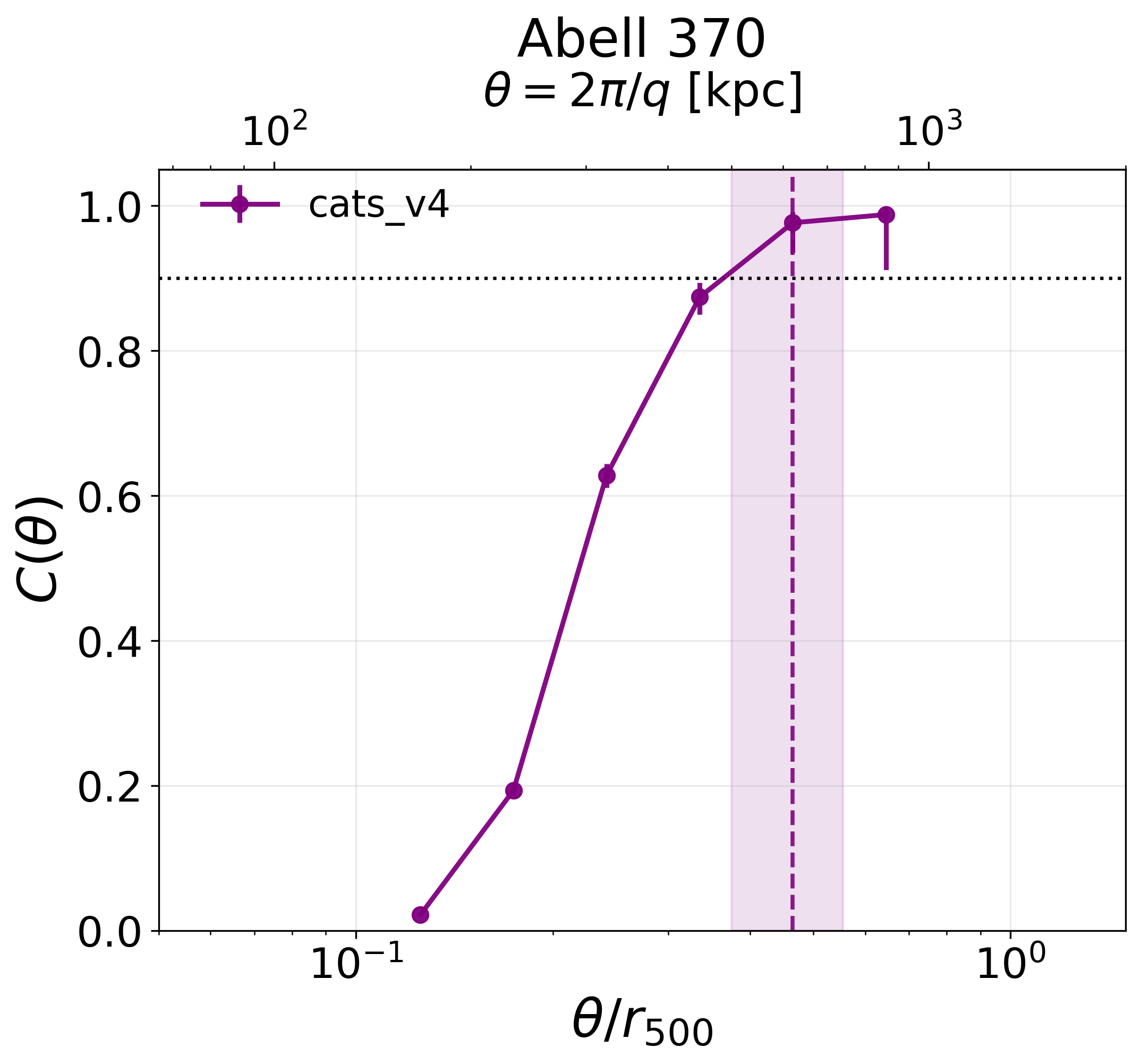}}
\subfigure[]{\includegraphics[width=0.28\textwidth]{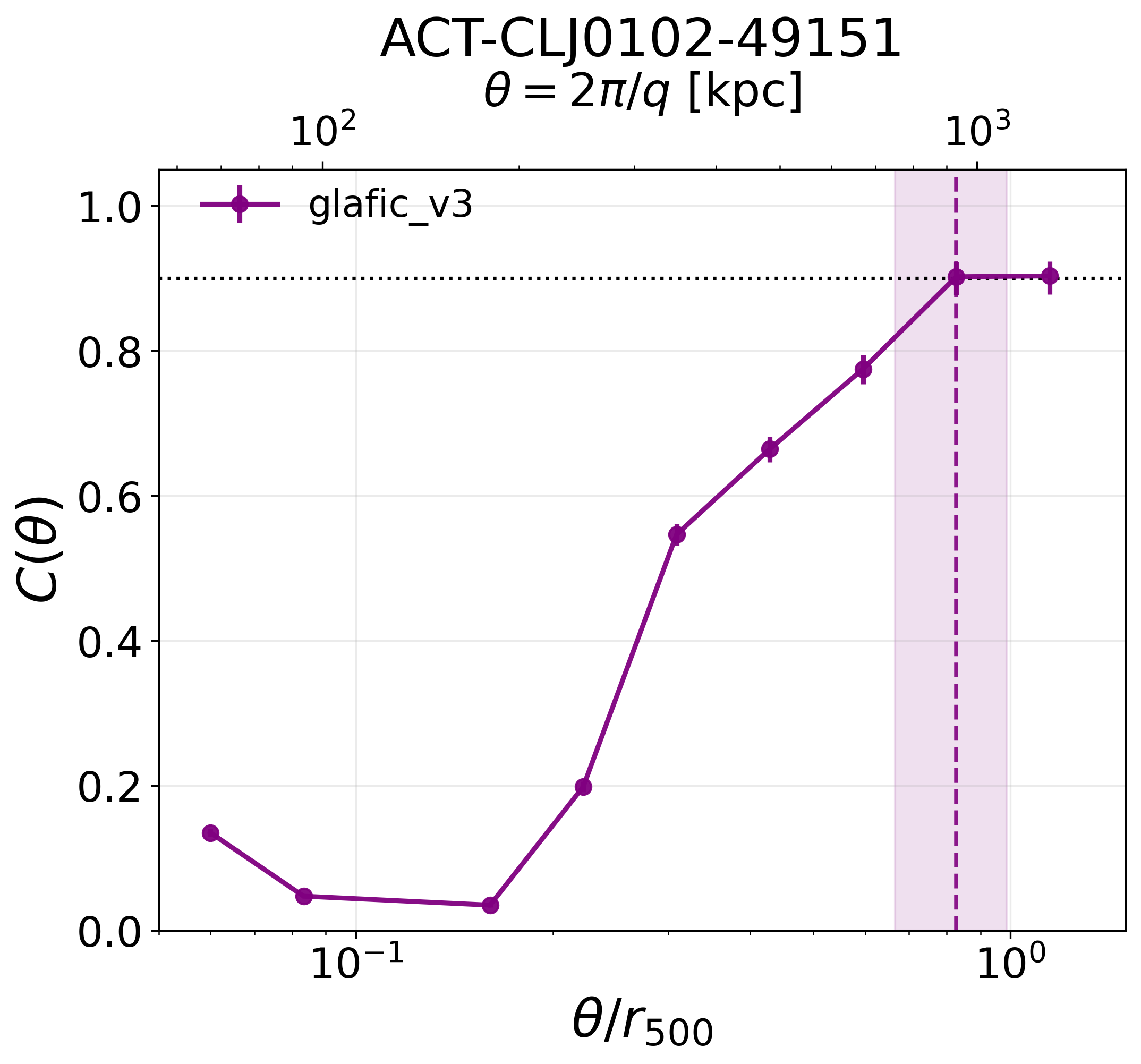}}\\

\subfigure[]{\includegraphics[width=0.28\textwidth]{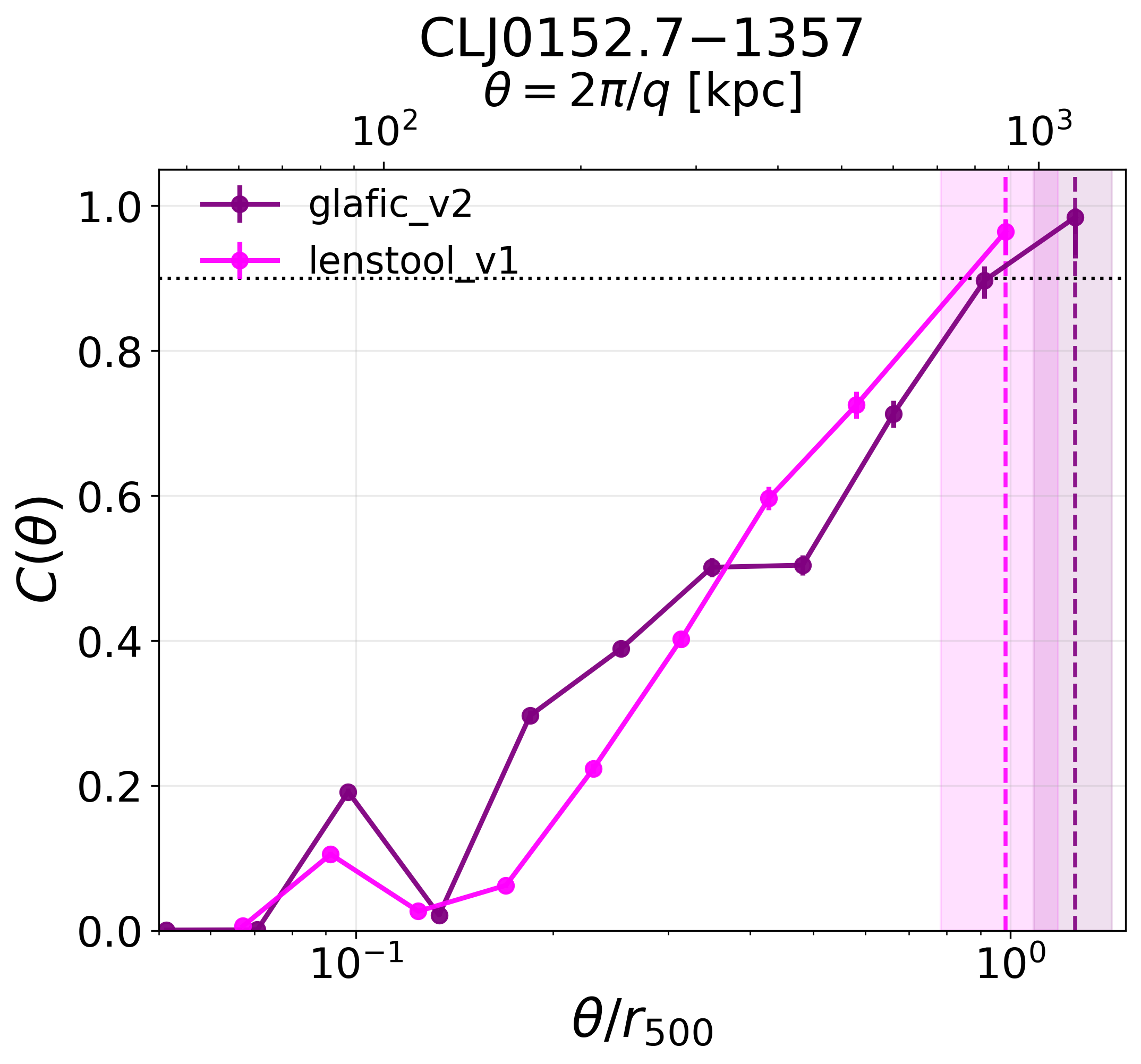}}
\subfigure[]{\includegraphics[width=0.28\textwidth]{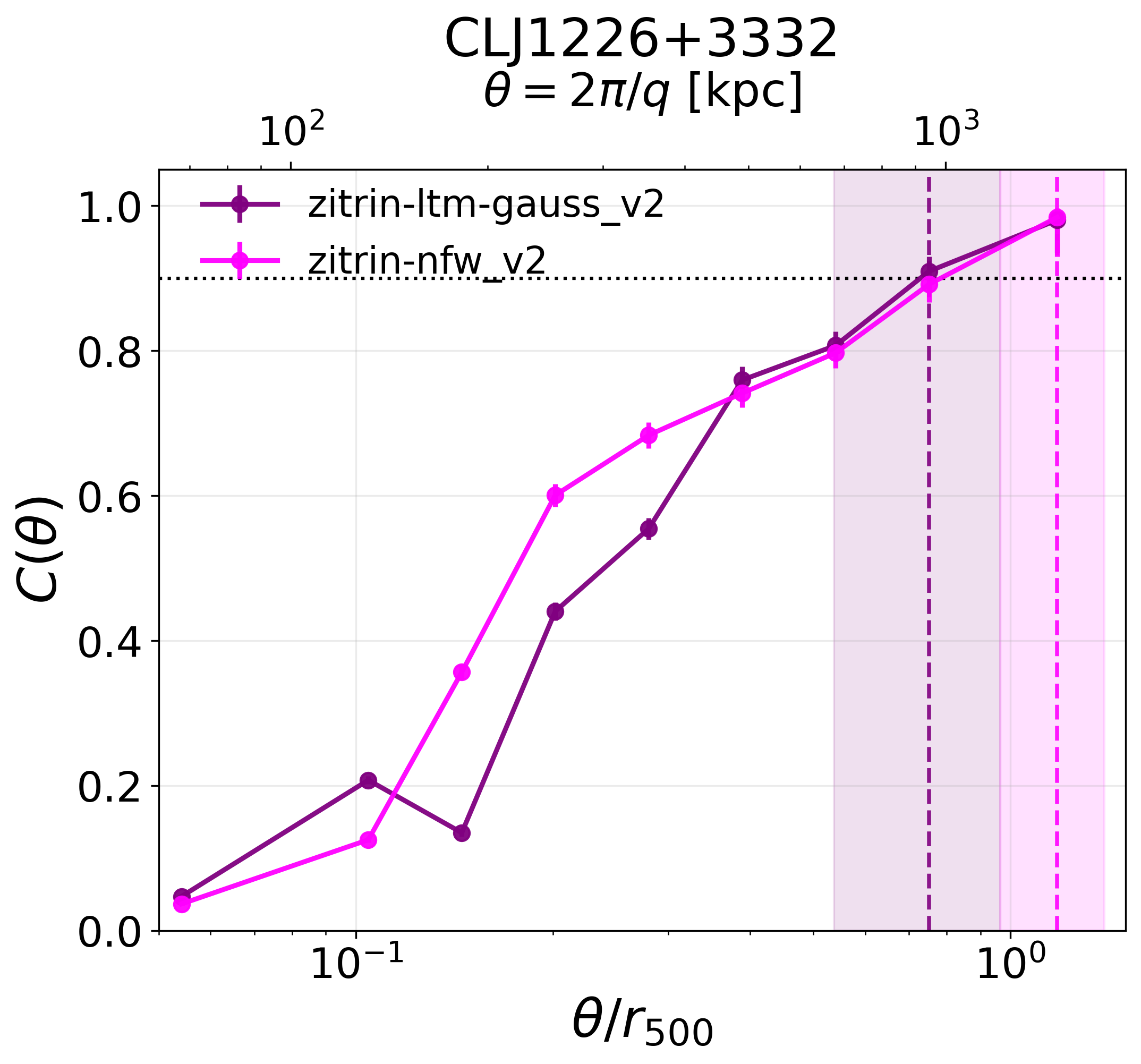}}
\subfigure[]{\includegraphics[width=0.28\textwidth]{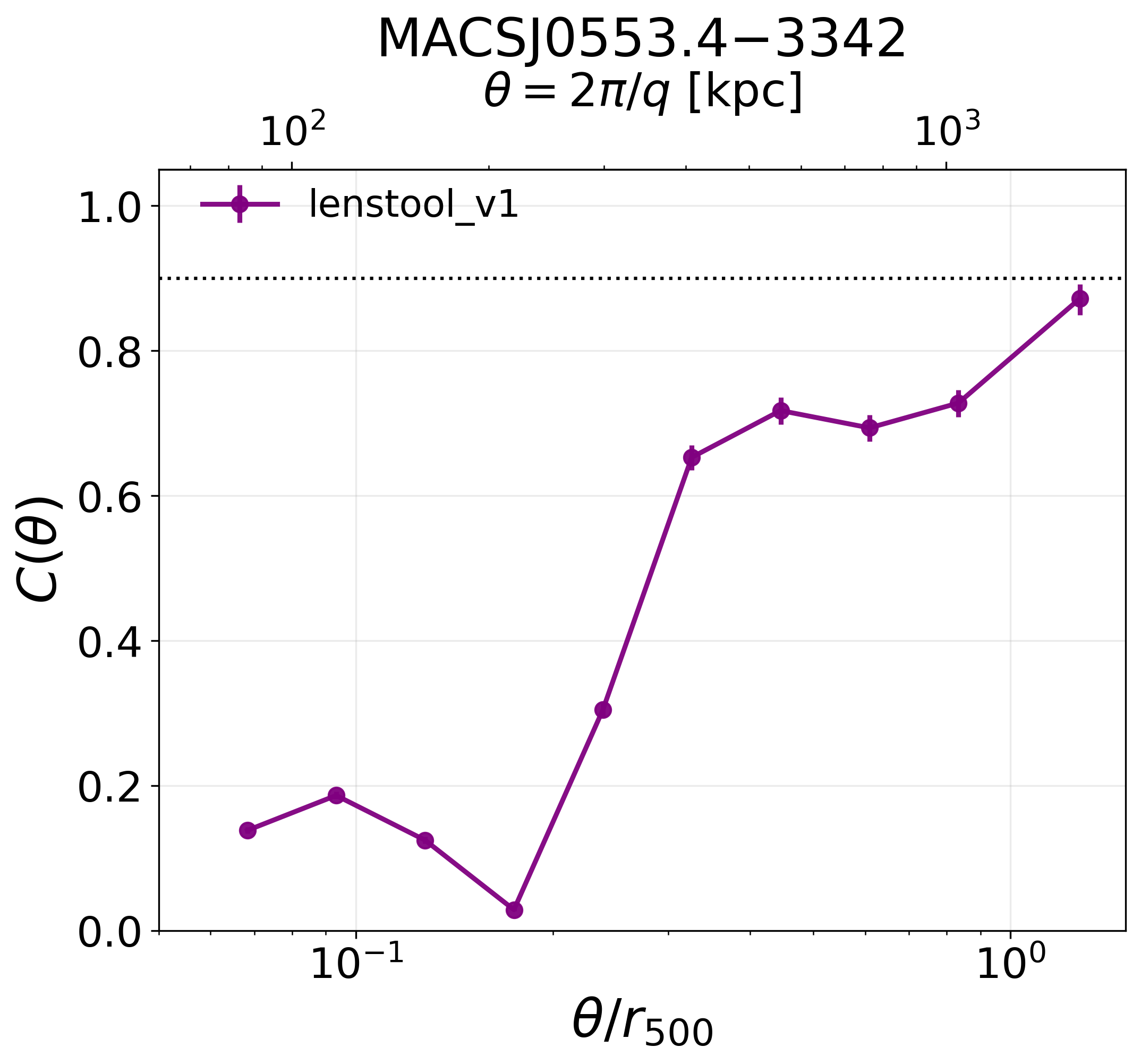}}\\

\subfigure[]{\includegraphics[width=0.28\textwidth]{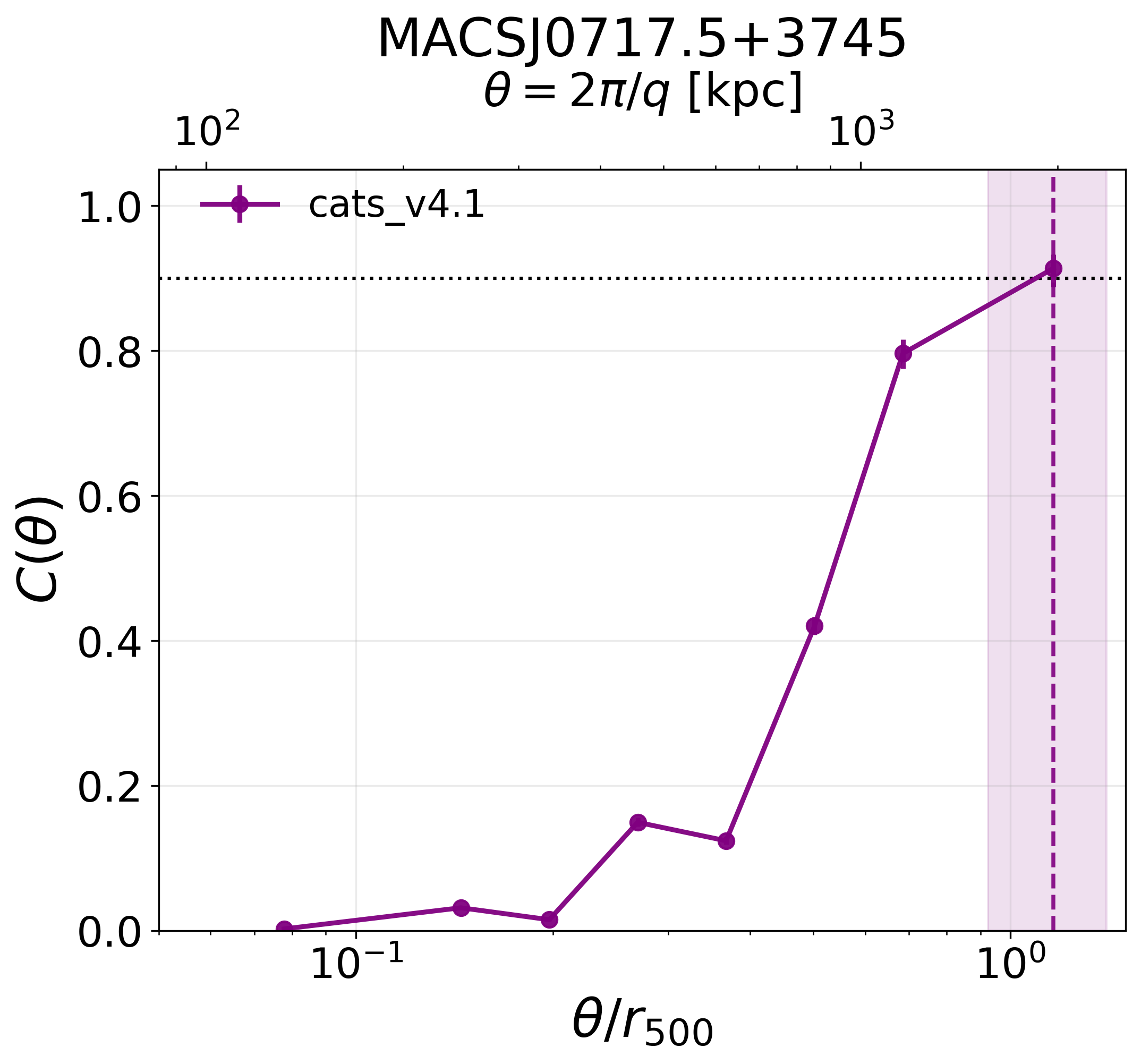}}
\subfigure[]{\includegraphics[width=0.28\textwidth]{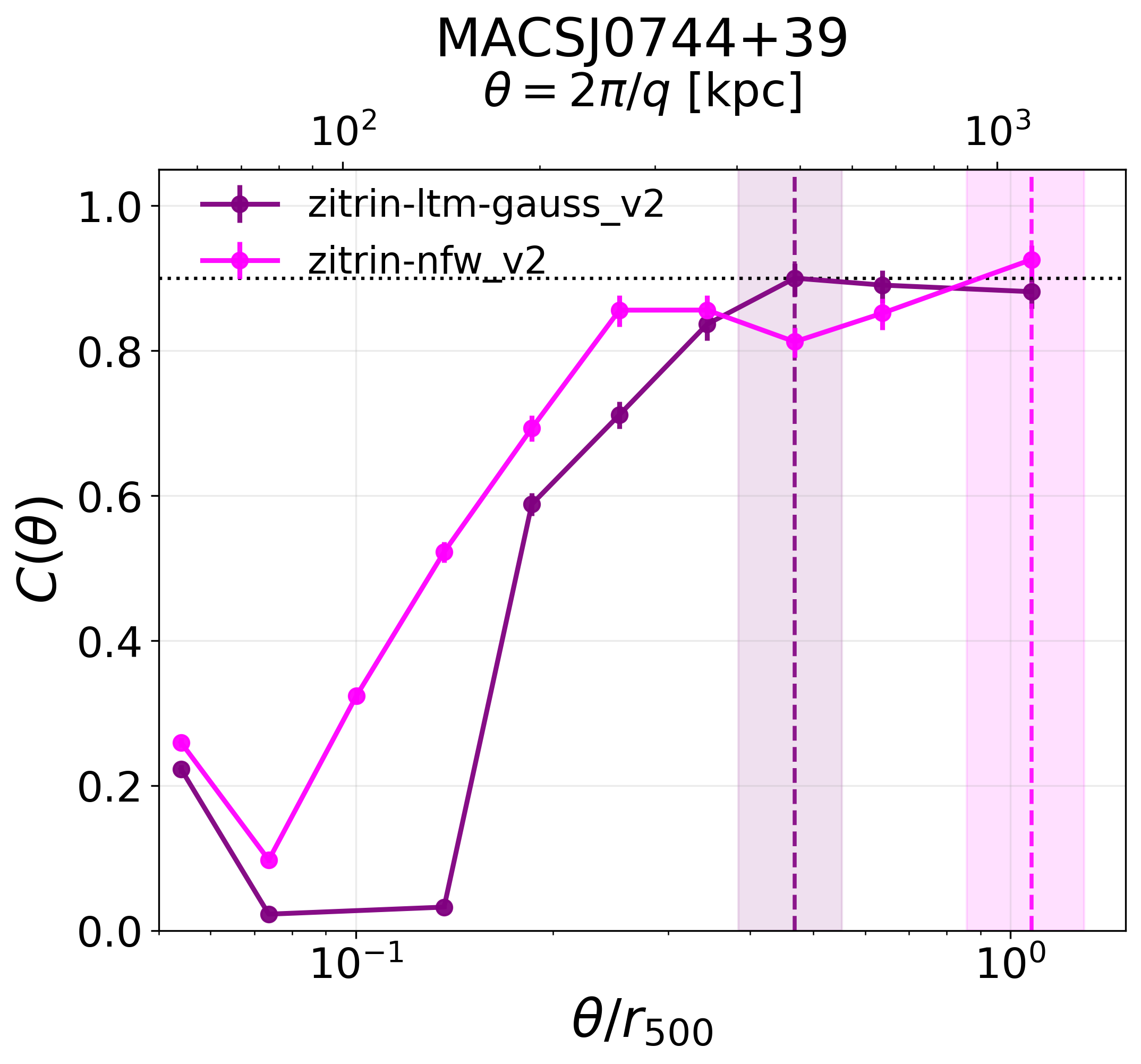}}
\subfigure[]{\includegraphics[width=0.28\textwidth]{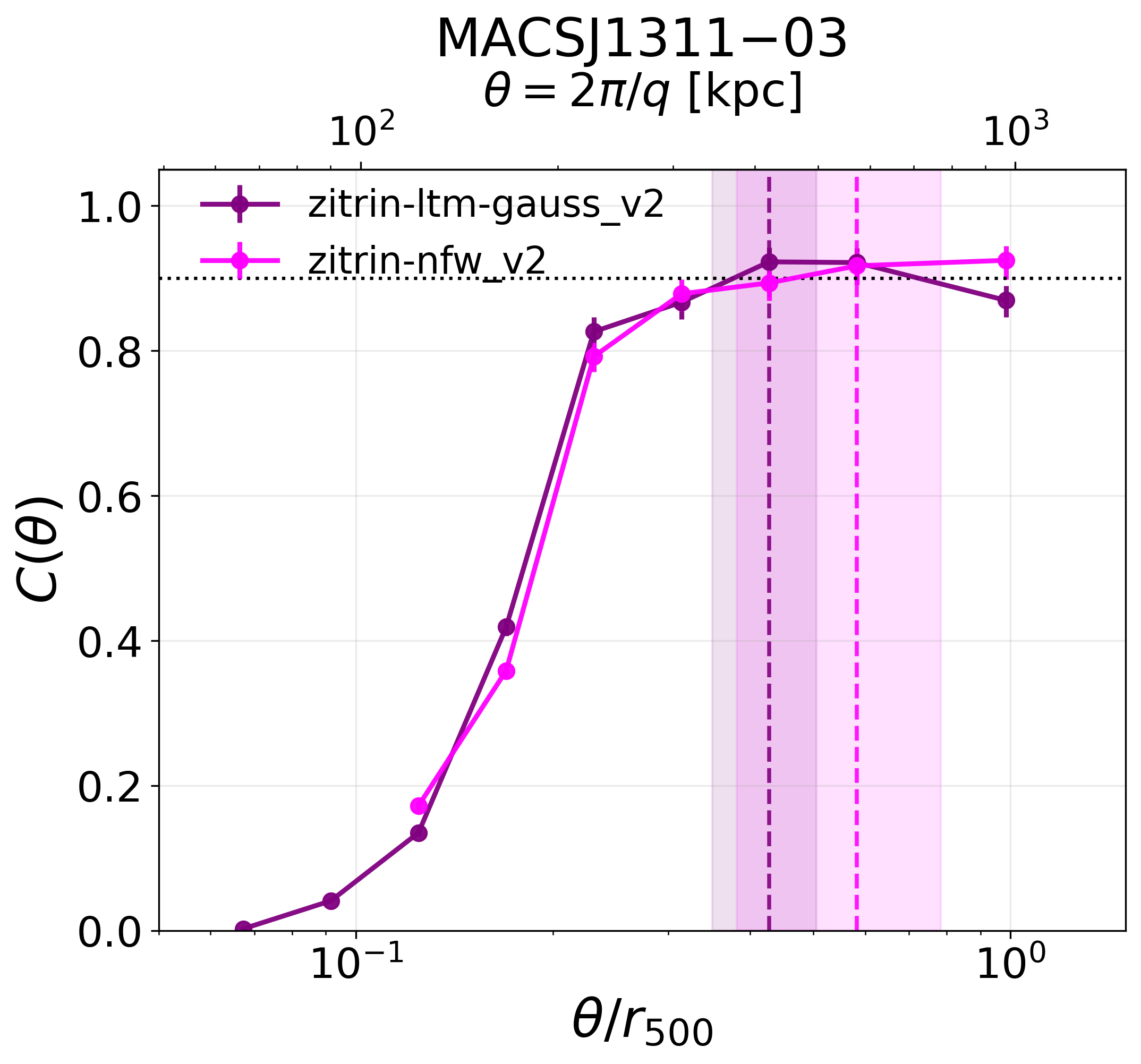}}\\

\subfigure[]{\includegraphics[width=0.28\textwidth]{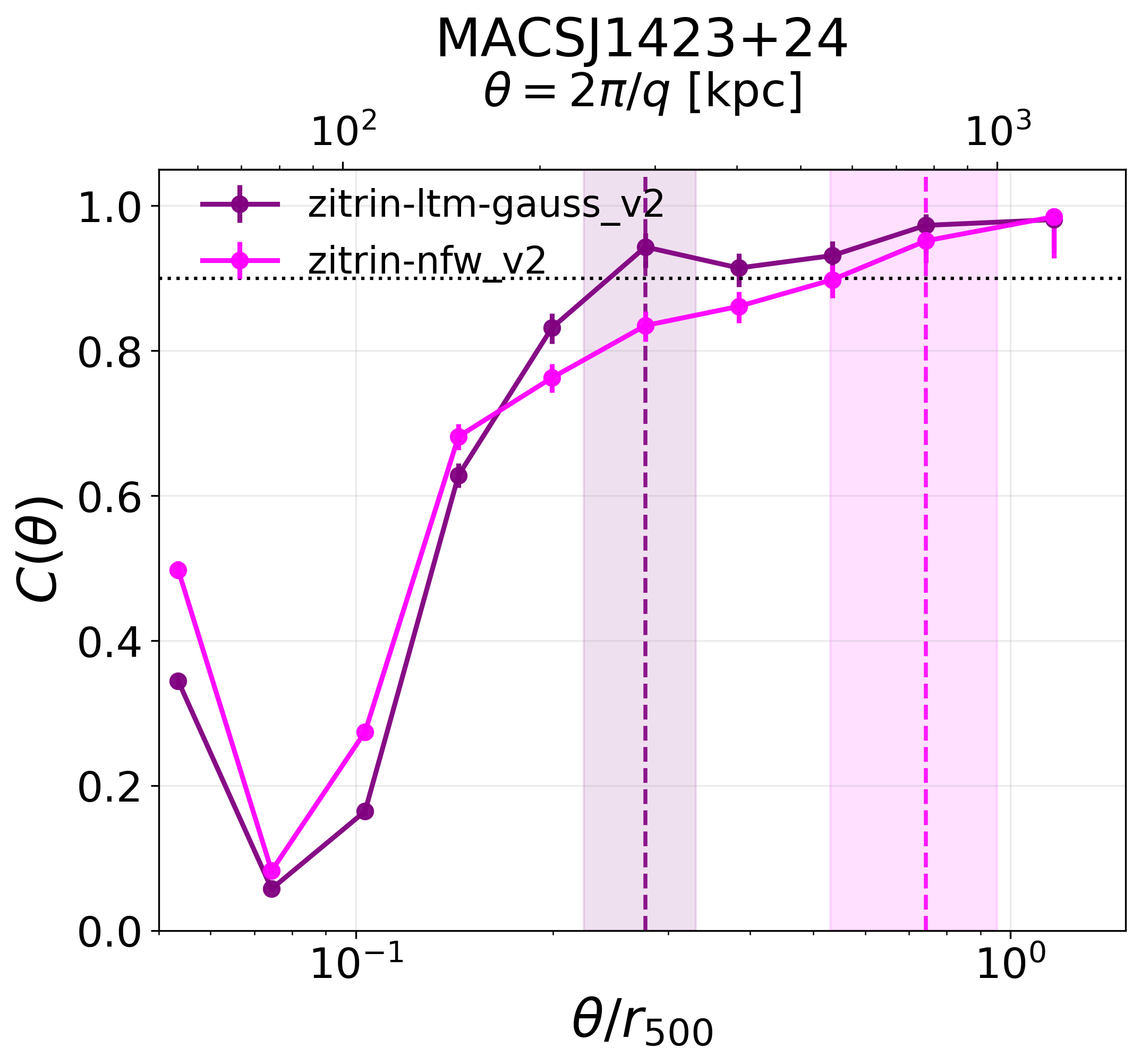}}
\subfigure[]{\includegraphics[width=0.28\textwidth]{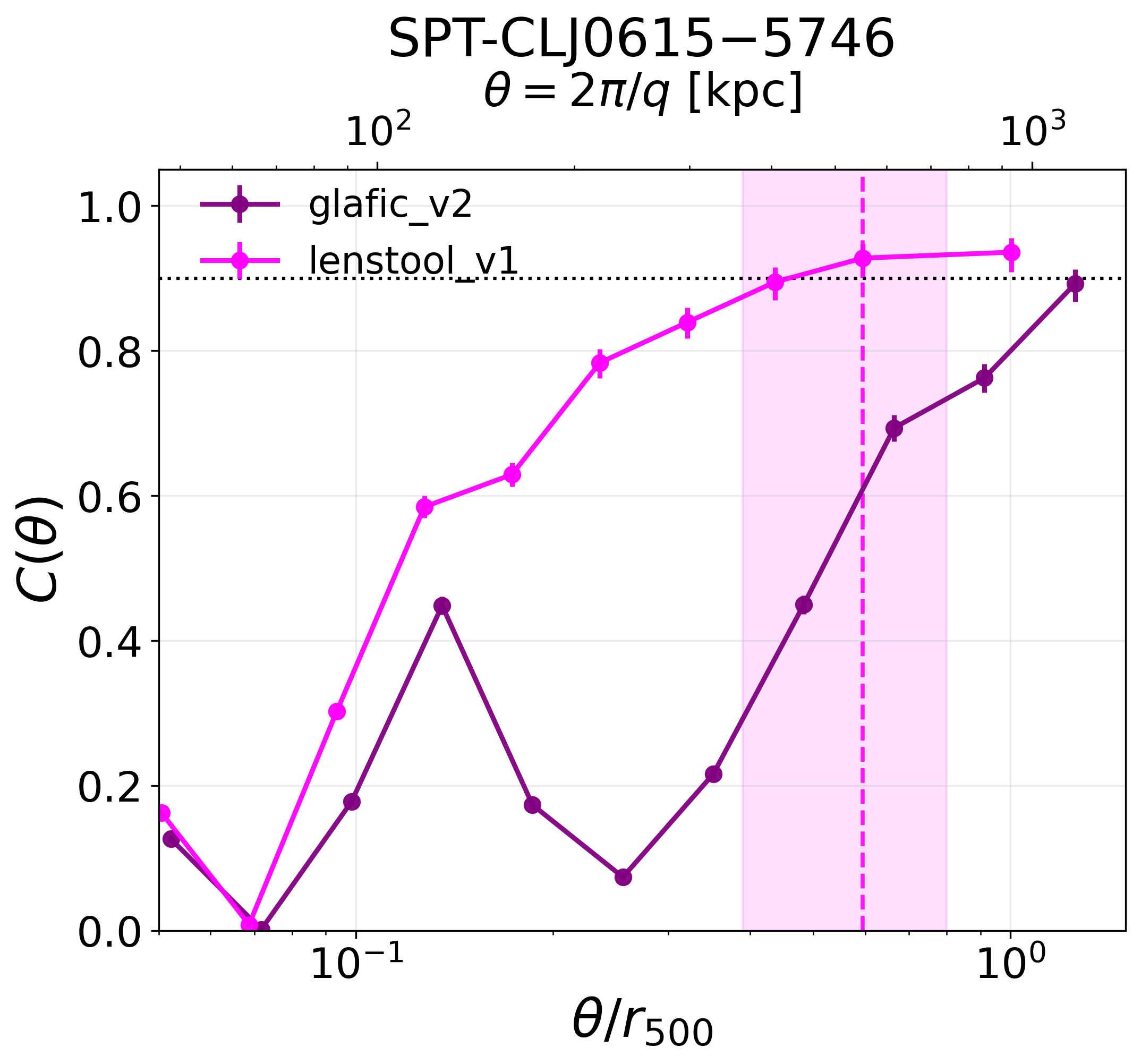}}

\caption{Coherence measurements for the subset of 11 clusters selected to satisfy the map-size criterion $L_{\kappa}^{\rm map}/(2r_{500}) > 0.8$, for which the inferred \textsl{coherence length} is expected to be more robust. Each panel shows the coherence as a function of scale (normalized by $r_{500}$) for all available lensing models of a given cluster, together with the corresponding \textsl{coherence length}.}
\label{reliable}
\end{figure*}

\textit{Software:} CIAO v4.17 \citealt{Fruscione2006}, HEASoft v6.33 \citealt{HEASoft}, Astropy v7.0.1 \citealt{Astropy2013,Astropy2018,Astropy2022}, pandas v2.1.4 \citealt{McKinney2010,pandas2020}, SciPy v1.15.2 \citealt{Virtanen2020}, Matplotlib v3.10.1 \citealt{Hunter2007}.

\section*{Acknowledgements}
This research has made use of data obtained from archival convergence maps from the HSTFF, CLASH, and RELICS programs. The HSTFF lensing mass maps are publicly available at \url{https://doi.org/10.17909/T9KK5N}, the CLASH lensing mass maps at \url{https://doi.org/10.17909/T90W2B}, and the RELICS lensing mass maps at \url{https://doi.org/10.17909/T9SP45}.

G.C., N.C., and P.N. acknowledge support from the NASA Astrophysics Data Analysis Program (ADAP) under grant \#80NSSC24K1025. P.N. also acknowledges P.N. also acknowledges support from the John Templeton Foundation via Grant \#126613. G.C. acknowledges the support of the Jet Propulsion Laboratory, California Institute of Technology, operated under contract with NASA (80NM0018D0004), where most of this work was carried out. G.C. also acknowledges Northeastern University for its support.

\newpage

%\end{acknowledgments}

%\appendix \label{appen}

\clearpage
\newpage

\bibliography{cluster_project}{}
\bibliographystyle{aasjournal}

\end{document}